\documentclass[12pt,a4paper]{article}
\pdfoutput=1
\usepackage{jheppub,tcolorbox}
\usepackage[utf8]{inputenc}
\usepackage{amsmath,amssymb}
\usepackage{float}
\usepackage{tikz} 
\usetikzlibrary{decorations.markings} 
\usetikzlibrary{decorations.text} 
\usetikzlibrary{positioning} 
\usetikzlibrary{backgrounds} 
\usetikzlibrary{fit} 
\usetikzlibrary{calc} 
\usetikzlibrary{shapes} 

\tikzset{dot/.style={draw,circle,inner sep=.7pt,fill,node
    distance=1cm}} 
\tikzset{dot1/.style={draw,circle,inner sep=.7pt,fill}} 
\tikzset{triangle/.style={draw,regular polygon, regular polygon
    sides=3}} 
\tikzset{->-/.style={decoration={
  markings,
  mark=at position .5 with {\arrow{>}}},postaction={decorate}}} 
\tikzset{-<-/.style={decoration={ 
  markings,
  mark=at position .5 with {\arrow{<}}},postaction={decorate}}}
\newcommand\be{\begin{equation}}
\newcommand\ee{\end{equation}}
\newcommand\bea{\begin{eqnarray}}
\newcommand\eea{\end{eqnarray}}

\begin{document}

\begin{titlepage}
\renewcommand{\thefootnote}{\fnsymbol{footnote}}

\vspace*{1.0cm}

\begin{center}
{\textbf{\huge Everything is a ``matter'' of perspective: the Unruh effect
}}
\end{center}
\vspace{1.0cm}

\centerline{
\textsc{\large J. A.  Rosabal}
\footnote{j.alejandro.rosabal@gmail.com}
}

\vspace{0.6cm}

\begin{center}
\it Asia Pacific Center for Theoretical Physics, Postech, Pohang 37673, Korea
\end{center}

\vspace*{1cm}

\centerline{\bf Abstract}

\begin{centerline}
\noindent

In this work we examine and extend the proposal of reference \cite{Rosabal:2018hkx},  concerning the new interpretation on the Unruh effect. The vacuum processes in Minkowski and Rindler space are described in detail,  in connection with the observers perspective. We highlight the presence of antiparticles in the radiation, which is perhaps the main observation in the cited reference. We present a new derivation of the vacuum energy in Rindler space using the Schr{\"o}dinger  kernel.  The quest for the Unruh radiation could be expanded by considering the possibility of detecting the antiparticles in it. A proposal for the experimental confirmation of the antiparticles in the radiation is presented.

\end{centerline}
\thispagestyle{empty}
\end{titlepage}

\setcounter{footnote}{0}

\tableofcontents

\newpage

\section{Introduction}\label{intro}

Recently it has been proposed a new way for visualizing the vacuum processes that take place in the Rindler wedge \cite{Rosabal:2018hkx}. These processes lead to the Unruh effect  \cite{Fulling:1972md,Davies:1974th,Unruh:1976db}.  The main prediction of \cite{Rosabal:2018hkx} is that the Unruh radiation is made of real particles but also  {\it real antiparticles}. New lines for the experimental confirmation of the effect
based on the presence of antiparticles in the radiation might be open.

In this work we extend the ideas presented in \cite{Rosabal:2018hkx}. We expand some of the concepts  and we present  a detailed  description of the perspective of the Minkowski and Rindler observer  about the vacuum processes. We pay special attention to the relation  between the quantum field theory (QFT) quantization   and the world line formalism in Rindler space. In addition a discussion about the experimental detection of the antiparticles in the radiation is presented. \\

One of the main predictions  in modern physics is perhaps the Hawking radiation \cite{Hawking:1974rv, Hawking:1974sw}. Despite, inspiring  remarkable ideas, as the holographic principle \cite{tHooft:1993dmi,Susskind:1994vu} and AdS/CFT  \cite{Maldacena:1997re,Witten:1998qj,Gubser:1998bc}, it lacks of experimental confirmation.   Moreover, it has generated several issues, the most famous is the information paradox \cite{Hawking:1976ra,Mathur:2009hf, Harlow:2014yka}. In the quest of the origin of the Hawking radiation and a solution for some of these paradoxes several interpretations have came up, some time generating even worse paradoxes, \cite{Hartle:1976tp,Susskind:1993if, Susskind:1993ws,Susskind:1994sm,Stephens:1993an,Kabat:1995eq,Almheiri:2012rt}. Unfortunately, the problems remain unsolved and the quest still goes on.

In principle, there could be  a possibility of cracking these problems, or at least learn more about them,  by studying an easier but conceptually similar phenomenon, the Unruh effect \cite{Fulling:1972md,Davies:1974th,Unruh:1976db}. This effect is intimately related to the Hawking radiation but the processes that lead to it take place in a flat space, where gravity does not play any role.\\

A Rindler observer (a non-inertial observer in Minkowski space having a proper constant acceleration)  \cite{Rindler:1966zz},  measures a vacuum energy $E_{vac}^R$,  given by the Plank distribution with a temperature
\be
T=\frac{\hbar \ {\bf a}}{2\pi\text{c} k},\label{temp-acele}
\ee
 proportional to the acceleration ${\bf a}$,
from now on we set $\hbar={\bf a}=\text{c}=k=1$.
In contrast, a Minkowski observer (an inertial observer with constant velocity),  measures a vanishing $E^{M}_{vac}=0$, vacuum energy.

The experimental confirmation of this effect,  or in other words, of the Unruh radiation, despite there is not gravity involved, could shed some light onto the origin of the Hawking radiation. There are several proposal for detecting this radiation, the most popular is the Unruh-DeWitt detector \cite{Unruh:1976db, DeWitt}. Unfortunately, neither by means of this detector nor by any other  means  the Unruh effect has been confirmed. In fact, nowadays,  there is a controversy around the origin of the particles in the radiation. Certainly,  this effect  has not been fully understood yet, see, for a review on this topic  \cite{CruzyCruz:2016kmi}   and references therein.\\

 In \cite{Rosabal:2018hkx}  a new interpretation to the Unruh effect was presented. This does not contradict  the one  proposed by Unruh \cite{Unruh:1976db}, it is somehow complementary to it. The new interpretation  highlights the presence of antiparticles in the radiation, and it was first reported in \cite{Rosabal:2018hkx}.

To extend the ideas of this reference here,  we will review  the quantization of  the massless free scalar field in
Rindler space using different boundary conditions to the ordinary ones in the canonical quantization, as presented in \cite{Rosabal:2018hkx}. The difference with the canonical quantization, where the initial value of the  field operator and its conjugate momentum are specified, is that  the field theory is quantized   imposing initial and  final conditions at the space-like slices of Rindler space that intersect  the points were the acceleration is turned on and off. This quantization makes manifest that a Rindler observer not necessarily has to be an eternally accelerated observer to obtain the Unruh effect.

We exploit the world line path integral representation of the propagator and the Schr{\"o}dinger kernel.  By using them in this  quantization scheme, i.e., with the boundary conditions we are using,  we can trace the vacuum virtual  particles running in loops in spacetime with
a Lorentzian metric. These loops give rise to the processes that contribute to the vacuum energy. It turns out that from the Rindler perspective some of these loops look as open paths. Precisely, these Rindler observer dependent open paths are the ones that contribute to the thermal radiation. In addition one can see that some of the open paths run backward in time, thus,  according to QFT this paths represent antiparticles.\\

The ordinary picture of the Unruh (Hawking) radiation is that  each particle is born entangled with an antiparticle. The former, living on the same region of the space where the Rindler observer lives, while the later,  living behind the horizon.
Here we find a similar picture, but in addition we find that  the Unruh (Hawking) radiation also contains entangled pairs of  particle-antiparticle, being both (the particle and the antiparticle) confined to  one side of the horizon. We believe that by using this fact  there would be more chance of confirming  the Unruh effect. These statements will be expanded and clarified  along the paper\\

The paper is organized as follows. In section \ref{vacproseccion} we describe the vacuum processes within QFT and the world line formalism. We also collect some of the ingredients  that we use throughout the paper. A brief description of Rindler space is presented. Section \ref{OPaTE} is devoted to introduce  the view of the Minkowski and Rindler observer, and what would be their interpretation of the vacuum processes. We present a picture which makes more clear the visualization of  the reality that each observer experiences. In the end we explain how particles and antiparticles are perceived by them. In section \ref{vacuumenergyseccion} we present a review of the vacuum energy calculation from the Rindler perspective  presented in \cite{Rosabal:2018hkx}. Experimental detection is discussed in section \ref{experimentdet}, emphasizing the presence of the antiparticles and its role in the detention of the Unruh radiation. We  also schematically present a possible experiment to confirm the effect. After conclusion,  appendix \ref{appA}, explaining the quantization in Minkowski and Rindler space, and a new  derivation, not reported so far in the literature,  of the vacuum energy in Rindler space using the Schr{\"o}dinger kernel, is presented.

\section{Vacuum processes}\label{vacproseccion}

\begin{figure}[hbt]
\centering
  \includegraphics[width=.5\textwidth]{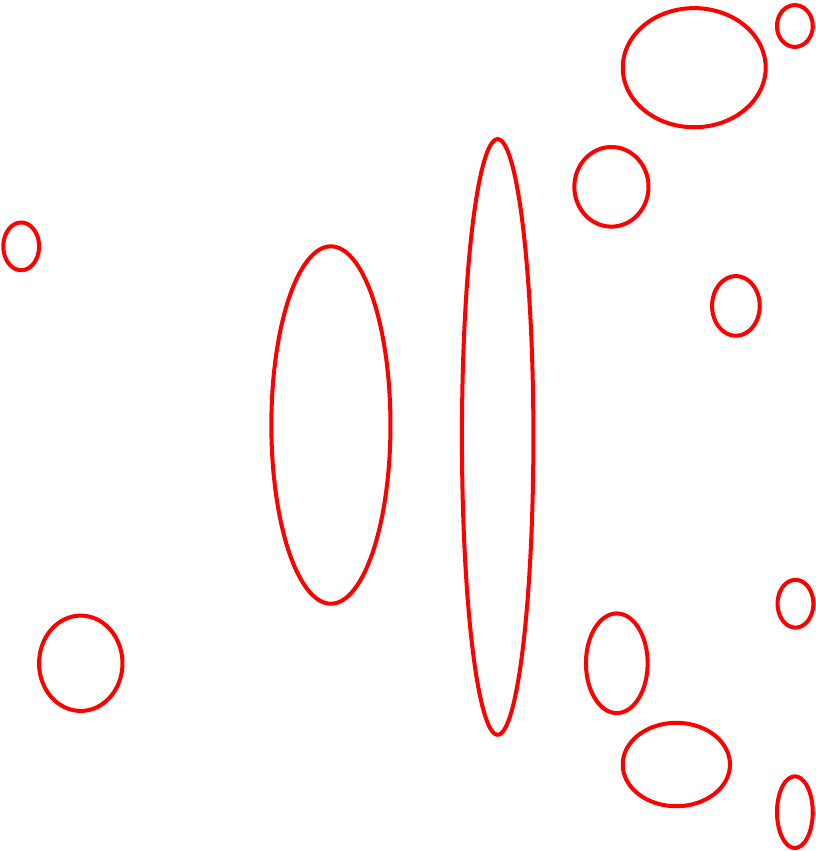}
  \caption{\sl Representation of the infinitely many  loops in Minkowski space that contribute  to the vacuum energy. They do not dependent of a particular observer.}\label{fig111}
\end{figure}

In this section we describe the vacuum processes in a free  QFT that contribute to the vacuum energy. In a non-interacting theory it is usually assumed that the only processes that contribute to the vacuum energy are the particles moving in  loops as  in Fig.\ref{fig111}. These processes take place in Minkowski space and they are a collection of events  that are independent of any observer (coordinate system).  We would like to stress that  each of the loops represented in Fig.\ref{fig111} is a representative of the infinitely many loops passing by some given points in spacetime (as they are treated in the world line formalism) and they do not represent the classical picture of several particles moving in loops.
\vspace{1cm}
\vspace{1cm}

In  QFT the vacuum energy can be written in terms of the one loop partition function. However,  the world line path integral  representation of the vacuum energy provides a more clear picture of the vacuum processes. In this formalism  the quantum particles can be interpreted  as moving over a collection of paths \cite{Feynman:1950ir,Strassler:1992zr,Bastianelli:2002fv,  Witten:2015mec, E.Witten}.

For example,  the generating functional $W\big[J\big]$ for a massless scalar field  in Minkowski space (with no boundaries)  is given by
\be\label{W}
W\big[J\big]=-\text{Tr}\text{Log}(\Box)+\text{i}\int d^dy_1\int d^dy_2J(y_2)(\Box)^{-1}_{y_2y_1}J(y_1).
\ee
For a massless  scalar field theory each term in \eqref{W} can be represented as a world line path integral \cite{Daikouji:1995dz}, as
\be
 -\text{Tr}\text{Log}(\Box) =\frac{1}{2}\int_0^{\infty}\frac{ds}{s}\int_{PBC}Dx^{\mu}(\tau)\text{exp}\Big[\text{i}\frac{1}{4s}\int_{0}^{1}d\tau\ \dot{\mathbf{x}}^2(\tau)\Big]\label{VE111},
\ee
and
\be
-\text{i}(\Box)^{-1}_{x_2x_1}=\int_0^{\infty}ds\int_{x(0)=x_1}^{x(1)=x_2}Dx^{\mu}(\tau)\text{exp}\Big[\text{i}\frac{1}{4s}\int_{0}^{1}d\tau\ \dot{\mathbf{x}}^2(\tau)\Big], \label{G111}
\ee
respectively,
where the functional integration for the first term is over all  loops in Minkowski space  and PBC stands for periodic boundary conditions. While for the second term the functional integration is over all  paths starting at $x_1$ and ending at $x_2$,  also in Minkowski space, and
\be
\dot{\mathbf{x}}^2(\tau)=-(\dot{x}^{0}(\tau))^2+(\dot{x}^{1}(\tau))^2+(\dot{x}^{2}(\tau))^2+(\dot{x}^{3}(\tau))^2.
\ee
Expressions \eqref{VE111} and \eqref{G111}  provide us with a geometrical view of  the quantum processes in Minkowski space, \cite{Feynman:1950ir}.

The one loop partition function $W\big[0]= -\text{Tr}\text{Log}(\Box)$,  is proportional to the  vacuum energy $E_{vac}$, see, for instance,  the discussion in p. 220 of \cite{Polchinski:1998rq},  and section \ref{OPaTE}. When a classical background  $J$  interacts with the field,   the  vacuum energy of the system, in this case  proportional to  $W\big[J\big]$,   will have two contributions as in \eqref{W}. One contribution coming from the loops \eqref{VE111} and one coming form the open path \eqref{G111}.

If the space has boundaries or horizons (and no classical background interacts with the field) the one loop partition function takes the form
\be\label{Wboundary}
W\big[\phi_b]=-\text{Tr}\text{Log}(\Box)+\text{i}\int d^{d-1}\sigma\sqrt{h(\sigma)}\phi_b(\sigma)\hat{n}^{\mu}\partial_{\mu}\phi_{cl} \big{|}_{\text{\tiny{boundary}}}.
\ee
Here, the $\text{Tr}$ is taken over the functions with vanishing boundary conditions, $\sigma$,  and $\sqrt{h}$ are coordinates and the measure associated to the induce metric on the boundary, respectively,  $\hat{n}^{\mu}\partial_{\mu}$,  is the normal derivative at the boundary and $\phi_{cl}$,  satisfies
\be
\Box \phi_{cl}=0 \quad;\quad \phi_{cl} \big{|}_{\text{\tiny{boundary}}}=\phi_b\label{equgeneral}.
\ee
Note that the general solution of \eqref{equgeneral} can be found using the Green's second identity,
\be
\phi_{cl}(y_2)=\int d^{d-1}\sigma_1\sqrt{h(\sigma_1)}\Big(\hat{n}^{\nu}\partial_{\nu}^{(1)}(\Box)_{y_2y_1}^{-1}\Big) \Big{|}_{y_1 \in \text{\tiny{boundary}}}\phi_b(\sigma_1)\label{generalgreensolution},
\ee
where without loss of generality $(\Box)_{y_2y_1}^{-1}$ can be chosen such that $(\Box)_{y_2y_1}^{-1} \big{|}_{y_1\in\text{\tiny{boundary}}}=0$.
Plugging \eqref{generalgreensolution} in \eqref{Wboundary} we get
\be
W\big[\phi_b]=-\text{Tr}\text{Log}(\Box)+\text{i}\int d^{d-1}\sigma_2\sqrt{h(\sigma_2)}\int d^{d-1}\sigma_1\sqrt{h(\sigma_1)}
\phi_b(\sigma_2)K(\sigma_2,\sigma_1)\phi_b(\sigma_1),\label{Wboundary1}
\ee
where
\be
K(\sigma_2,\sigma_1)=\Big[\hat{n}^{\mu}\partial_{\mu}^{(2)}\Big(\hat{n}^{\nu}\partial_{\nu}^{(1)}(\Box)_{y_2y_1}^{-1}\Big)\Big]\Big{|}_{(y_1,y_2)\in\text{\tiny{boundary}}},
\ee
can be identified  with the boundary-to-boundary propagator \cite{Witten:1998qj,Kabat:1995eq,Rehren:2004yu}.
Note that \eqref{W} and \eqref{Wboundary1} are structurally similar. Also note that the boundary value of the field $\phi_b(\sigma)$, in \eqref{Wboundary1}  plays the role of the source $J$, in \eqref{W}. In fact, if \eqref{Wboundary1} were regarded as a generating functional the derivatives with respect to the boundary values of the field $\phi_b(\sigma)$ would lead to the  boundary-to-boundary propagator. Conceptually, however, they are quite different. While the former contains open path contributions connecting two points which could be everywhere in the space, the later contains open path contributions connecting points starting and ending only at the boundaries or horizons.

For Rindler space, for example, where there are two horizons, one at past infinity and the other at future infinity,  in the Rindler time, we will have to specify two boundaries  values in order to fully determine \eqref{Wboundary}. More generically if the Rindler system is bounded by two space-like  slices $\tau_i$ and $\tau_f$, \eqref{Wboundary} in $d=4$, will have the form
\be\label{WRindler}
W\big[\phi_i,\phi_f]=-\text{Tr}\text{Log}(\Box_R)+\text{i}\int \frac{d\rho}{\rho} \  d^{2}y \Big(\phi_f\partial_{\tau}\phi_{cl}\big{|}_{\tau_f}-\phi_i\partial_{\tau}\phi_{cl}\big{|}_{\tau_i}\Big).
\ee
For this case $\Box_R \phi_{cl}=0$, with the Laplacian $\Box_R$,  in Rindler space and boundary conditions
\bea\label{bounRin}
\phi_{cl}\big{|}_{\tau_i} & = & \phi_i, \nonumber \\
\phi_{cl}\big{|}_{\tau_f}  & = & \phi_f.
\eea
As mentioned, in this case,  the vacuum energy of the system  is proportional to  $W\big[\phi_i,\phi_f]$,
\be
E^R_{vac}\propto W\big[\phi_i,\phi_f],
\ee
 and it will have two contributions. One coming form the loops inside $\tau_i<\tau<\tau_f$,  and a second one coming from the open path starting and ending at the boundaries (horizons).  This picture, that will be clarified in the subsequent sections,  resembles the holography setup. One could use the same techniques as in holography \cite{Witten:1998qj, Gubser:1998bc}, say compute the kernel (Green function) for solving $\Box \phi_{cl}=0$ with the boundary conditions \eqref{bounRin}, explicitly compute $W\big[\phi_i,\phi_f]$ and then relate the result with \eqref{VE111}  and \eqref{G111}. In this paper however, we will use the operator formalism to compute $E^R_{vac}$.

\subsection{Rindler  space}

In this section we present a brief description of Rindler  space \cite{Rindler:1966zz}, or right Rindler wedge, $M_R$. It is defined as the region $x^1>|x^0|$, in a two-dimensional Minkowski space. It can be extended to higher dimensions. In four dimensions, the $M_R$ metric can be written as
\be
ds^2  =    -\rho^2 d\tau^2+d\rho^2+(dx^2)^2+(dx^3)^2  \label{metric1}.
\ee
It is related  to the flat metric
\be
ds^2 = -(dx^0)^2+(dx^1)^2+(dx^2)^2+(dx^3)^2,
\ee
by the coordinate transformation
\bea\label{transcoor111}
x^0 (\tau,\rho) & = & \rho \ \text{sinh}(\tau), \nonumber\\
x^1(\tau,\rho) & = & \rho \ \text{cosh}(\tau).
\eea

\begin{figure}[]
\centering
  \includegraphics[width=.4\textwidth]{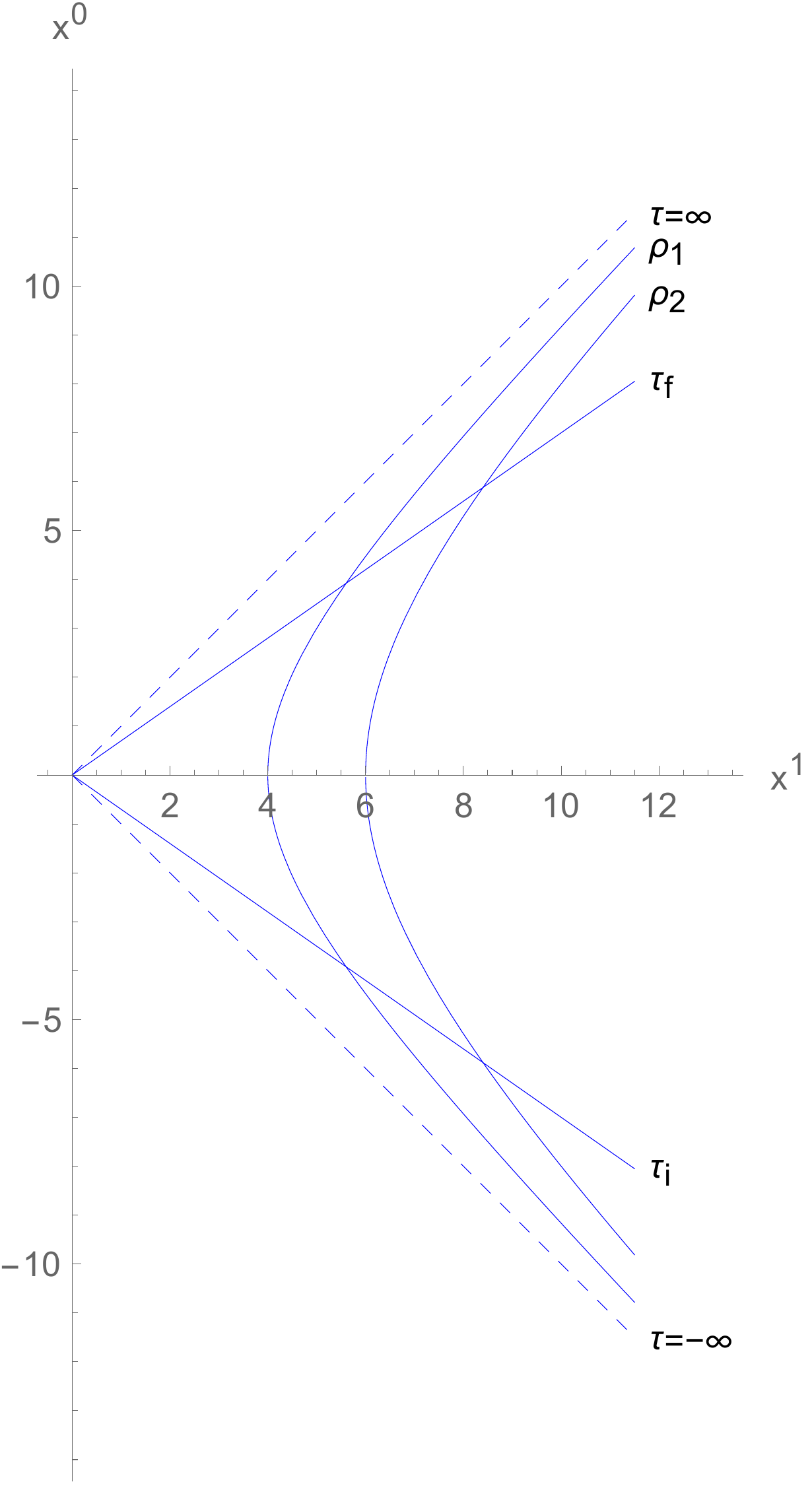}
  \caption{\sl Rindler slicing,  $\tau={-\infty}<\tau_i<\tau_f<\tau={\infty}$, and $\rho_1<\rho_2$ .} \label{slicing}
\end{figure}

The foliation of the Rindler wedge can be obtained as follows. We can start with  (\ref{transcoor111}) and eliminate the $\tau$ dependence, i.e.,
\be
\big(x^1(\tau,\rho) \big)^2-\big(x^0 (\tau,\rho)\big)^2=\rho^2.
\ee
It means that  each  constant $\rho$ represents a hyperbola. On the other hand, eliminating the $\rho$ dependence,
\be
x^0 (\tau,\rho) =\text{tanh}(\tau)x^1 (\tau,\rho),
\ee
we see that each constant $\tau$ represents a straight  line starting at the origin. Notice that when $\tau\rightarrow\pm\infty$, $x^0 =\pm x^1$, which are the Rindler horizons.
The Rindler  slicing is represented in Fig.\ref{slicing}.

\section{Observers perspective}\label{OPaTE}

\begin{figure}[H]
\centering
  \includegraphics[width=.5\textwidth]{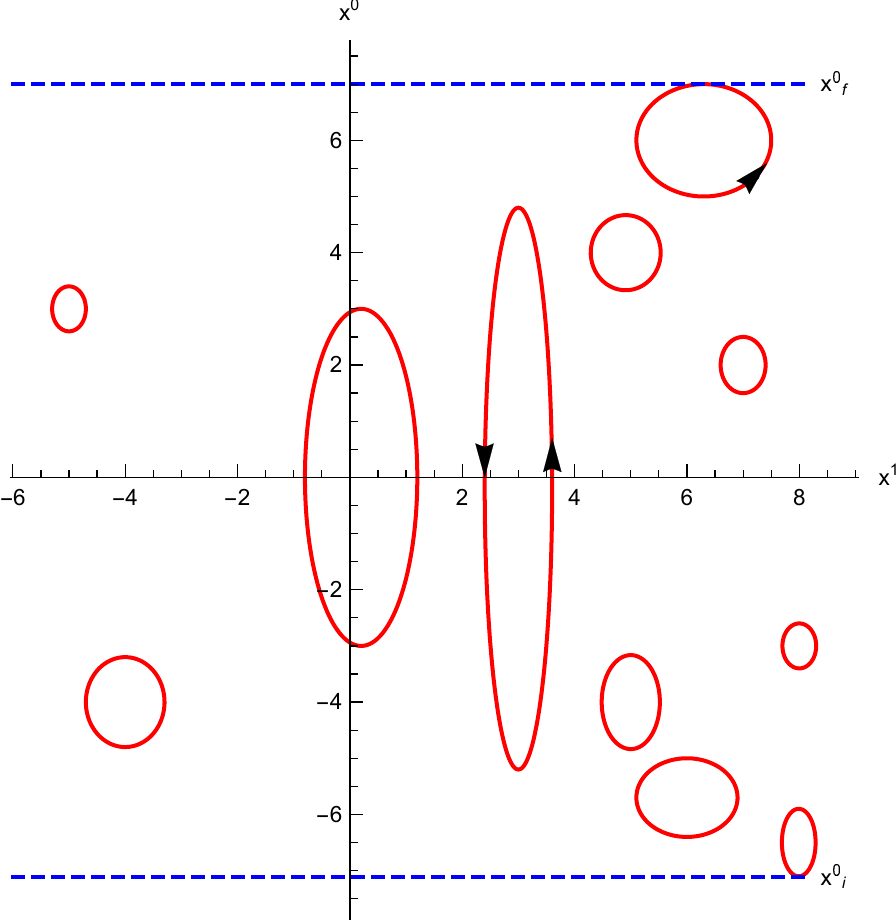}
  \caption{\sl Representation of the inertial observer perspective  of the infinitely many  loops in Minkowski space  relevant to the vacuum energy calculation.  The arrows represent the flow of time in the loop.} \label{fig222}
\end{figure}

In order to create a picture of  how each observer ``sees'' the vacuum processes, we will present a detailed description of them based on the world line path integral formulation of QFT. Let us start asking, how a  Minkowski observer would compute the vacuum energy using the world line formalism. This observer will consider all the loops in  Minkowski space, Fig.\ref{fig222}, with $x^0_i\rightarrow-\infty$ and $x^0_f\rightarrow+\infty$.

 For establishing the relation between the loops and the vacuum energy $E^{M}_{vac}= E^{M}_{0}$ in Minkowski space  we will work, for simplicity,  with  a massless non-interacting  scalar QFT,  see appendix  \ref{appA}. The results however, can be extended to others fields.  The vacuum energy is given by
 \be
 E^{M}_{0}=\frac{1}{2}\delta^3(0)\int_{-\infty}^{\infty}d^3p \  |p|,\label{e0minkprimera}
  \ee
  where $|p|=\sqrt{p_1^2+p_2^2+p_3^2}$.
Using
\be
\delta^3(k)=\frac{1}{(2\pi)^3}\int_{-\infty}^{\infty}d^3x \ \text{e}^{\text{i}k\cdot x},
\ee
which leads to
\be
\delta^3(0)=\frac{1}{(2\pi)^3}\int_{-\infty}^{\infty}d^3x=\frac{V_3}{(2\pi)^3}, \label{volumendelta}
\ee
we can express  the vacuum energy as
 \be
 E^{M}_{0}=\frac{1}{2}V_3\int_{-\infty}^{\infty}\frac{d^3p}{(2\pi)^3} \  |p|.
  \ee
Now we can use the regularized integral (see  p. 220 of \cite{Polchinski:1998rq})
\be
\text{i}\int_{0}^{\infty}\frac{ds}{s}\int_{-\infty}^{\infty}\frac{dp_0}{2\pi}\text{e}^{\text{i}s (-p^2_0+|p|^2)}= |p| , \label{importantidentity}
\ee
to rewrite $E^{M}_{0}$ as
\be\label{ecerominko}
 E^{M}_{0}=\text{i}\frac{1}{2}V_3\int_{0}^{\infty}\frac{ds}{s}\int_{-\infty}^{\infty}\frac{d^4p}{(2\pi)^4}\text{e}^{\text{i}s (-p^2_0+|p|^2)}.
  \ee
Using the identity
\be
\int_{-\infty}^{\infty}\frac{d^4p}{(2\pi)^4}\text{e}^{\text{i}s(-p^2_0+|p|^2)}=\frac{1}{V_4}\int_{PBC}Dx(\tau)Dp(\tau)\text{e}^{\big[-\text{i}\int_{0}^{1}d\tau\big( p_{\mu}(\tau)\dot{x}^{\mu}(\tau)-s \eta^{\mu\nu}p_{\mu}(\tau)p_{\nu}(\tau) \big)\big]}.\label{identidadinteg}
\ee
It can be easily derived  \cite{Gordon:2014aba} by integrating \eqref{identidadinteg} in $x(\tau)$. Then integrating \eqref{identidadinteg}  in $p(\tau)$ we get the path integral representation of the vacuum energy.
\be
 E^{M}_{0}=\text{i}\frac{1}{2}\frac{1}{V_1}\int_{0}^{\infty}\frac{ds}{s}\int_{PBC}Dx(\tau)\text{e}^{\big[\text{i}\frac{1}{4s}\int_{0}^{1}d\tau \ \eta_{\mu\nu}\dot{x}^{\mu}(\tau)\dot{x}^{\nu}(\tau)\big]}, \qquad V_1=\int_{-\infty}^{\infty}dx^0\label{e0M},
 \ee
with $\eta_{\mu\nu}=(-1,1,1,1)$. Having shown that the vacuum energy is related to the loop integration in the world line path integral, it is worth to remark that  when sufficient care is taken \cite{Feynman:1950ir}  the path integration can be performed over all the loops in  Minkowski space.

 A Rindler observer would proceed in a similar way. The  perspective of this observer is restricted to the wedge. Then, according to the prescription that only the loops will contribute to the vacuum energy  $E^{R}_{0}$, this observer will regard all the loops in $M_R$.

 Now, we shall present  the details of what would be the (naive) calculation of $E^{R}_{vac}=E^{R}_{0}$,  from  the Rindler observer point of view.  In analogy with the Minkowski observer, a Rindler observer would use the analogous expression to \eqref{e0M} but in Rindler space
 \bea\label{pathintRindler1}
 E^{R}_{0} & = & \text{i}\frac{1}{2}\frac{1}{V_1}\int_{0}^{\infty}\frac{ds}{s}\int_{PBC}Dy(\sigma)\sqrt{-g(y(\sigma))}\text{e}^{\big[\text{i}\frac{1}{4s}\int_{0}^{1}d\sigma \ g_{\mu\nu}(y(\sigma))\dot{y}^{\mu}(\sigma)\dot{y}^{\nu}(\sigma)\big]}\\
{} & = &   \text{i}\frac{1}{2}\frac{1}{V_1}\int_{0}^{\infty}\frac{ds}{s}\int_{\tau_i}^{\tau_f}d\tau \int_0^{\infty} \frac{d\rho}{\rho}\int d^2\bar{y} K_R(y,y,s) \label{pathintRindler0},
 \eea
where $y^{\mu}=(\tau,\rho,x^2,x^3)$, $\sigma$ is the proper time, $K_R(y,y^{\prime},s)$ is the Schr{\"o}dinger kernel that satisfies the equation\footnote{Usually for this kind of manipulations one uses the Wick rotated version of the Schr{\"o}dinger kernel which is the heat kernel. This is to ensure the convergence of the integrals involved in the calculations.}
 \be
 \text{i}\partial_{s}K_R(y,y^{\prime},s)=-\frac{1}{2}\hat{O} K_R(y,y^{\prime},s), \label{Schrodinger_kernel }
 \ee
 and for this discussion we will take the metric (\ref{metric1}). We have to proceed with caution when using this method for computing the Schr{\"o}dinger kernel  in this space. The operator that appears on the right hand side of \eqref{Schrodinger_kernel }, strictly speaking, is not the Laplacian in Rindler space. It is given by
\be
\hat{O}=\partial_{\tau}\partial_{\tau}-\rho\partial_{\rho}(\rho\partial_{\rho})-\rho^2\delta^{ab}\partial_a\partial_b\label{apeempa}.
\ee
We refer to \cite{Emparan:1994qa} for a discussion on the heat kernel on Rindler space and the subtleties with the operator \eqref{apeempa} and expression \eqref{pathintRindler0}.

The Rindler wedge does not cover the whole Minkowski space, so the path integral (\ref{pathintRindler1}) is over all the loops inside the Rindler wedge, which is all the space this observer experiences. Solving (\ref{pathintRindler1}) is a difficult task. Nevertheless, we can take advantage of the fact that equation (\ref{Schrodinger_kernel }) can be easily solved. Similarly to the Minkowski calculation in  appendix \ref{appA}, regarding all the loops in Rindler space translates into the only boundary condition $K_R(y,y^{\prime},s)\big{|}_{s\rightarrow \pm\infty} =  0$.
The Schr{\"o}dinger kernel can be expressed as
 \be
 K_R(y,y^{\prime},s)=\sum_{n}\chi_n(y)\chi_n^{*}(y^{\prime})\text{e}^{\text{i}\lambda_n\frac{s}{2}},
 \ee
 where $\chi_n(y)$ are the normalised eigenfunctions  and $\lambda_n$ are the eigenvalues of the operator $-\frac{1}{2}\hat{O}$. For Rindler space we have,
 \be
K_R(y,y^{\prime},s)= \int_{-\infty}^{\infty}\frac{d\nu_0}{2\pi} \int_{0}^{\infty}d\nu \int_{-\infty}^{\infty}\frac{d^2k}{(2\pi)^2}\text{e}^{\text{i}\nu_0(\tau-\tau^{\prime})} \psi_{\nu,\bar{k}}(\rho)\psi_{\nu,\bar{k}}(\rho^{\prime})\text{e}^{\text{i}\bar{k}\cdot(\bar{x}-\bar{x}^{\prime})} \text{e}^{\text{i}(-\nu_0^2+\nu^2)\frac{s}{2}}.
 \ee
The details and some properties of the functions $\psi_{\nu,\bar{k}}(\rho)$, can be found  in the next section as well as in appendix \ref{appA}.

Plugging  the kernel in (\ref{pathintRindler0}) yields to
 \be
 E^{R}_{0} = \text{i}\frac{1}{2}\frac{V_3}{V_1}\delta(0)\int_{0}^{\infty}\frac{ds}{s} \int_{-\infty}^{\infty}\frac{d\nu_0}{2\pi} \int_{0}^{\infty}d\nu \int_{-\infty}^{\infty}\frac{d^2k}{(2\pi)^2} \text{e}^{\text{i}(-\nu_0^2+\nu^2)\frac{s}{2}},
 \ee
 where $\delta(0)$, comes from the orthogonality relation between the $\psi_{\nu,\kappa}(\rho)$, functions
\be
\int\limits_0^{\infty}\frac{d\rho}{\rho}\ \psi_{\nu,\kappa}(\rho)\psi_{\nu^{\prime},\kappa}(\rho)=\delta(\nu-\nu^{\prime}).
\ee
 Using (\ref{importantidentity}) and $\delta(0)=\frac{V_1}{2\pi}$, we get
 \be
 E^{R}_{0} = V^{R}_{3} \int_{-\infty}^{\infty}\frac{d^2k}{(2\pi)^2}  \int_{0}^{\infty}\frac{d\nu}{2\pi} \nu=\frac{1}{2}\delta(0)^3 \int_{-\infty}^{\infty}d^2k  \int_{0}^{\infty}d\nu \nu\label{verw},
 \ee
 where $V_3^R=\frac{1}{2}V_3$, is the volume of Rindler space.

We have computed the vacuum energy in the Rindler wedge and we have not obtained  the thermal distribution, as expected for the Unruh effect. What is wrong in the previous calculation? In fact, there is nothing wrong, what happens is that  this (naive) calculation is incomplete because of there are contributions that  we are not taking into account. In other words, the Schr{\"o}dinger kernel must satisfy further  boundary conditions. In appendix \ref{appA} we derive the proper Schr{\"o}dinger kernel in Rindler space. In what follows, however,  we want to proceed in a more intuitive fashion. We would like to see whether the vacuum energy in Rindler space really contains other contributions coming from the open paths, as stated in the discussion involving equation \eqref{WRindler},  i.e.,  could it be the case that
\be
E^R_{vac}=E^R_{\text{(open paths)}}+E^R_{0}\ ?\label{rindlerenergyopenpaths}
\ee

\begin{figure}[ht!]
 \centering
   \includegraphics[width=.5\textwidth]{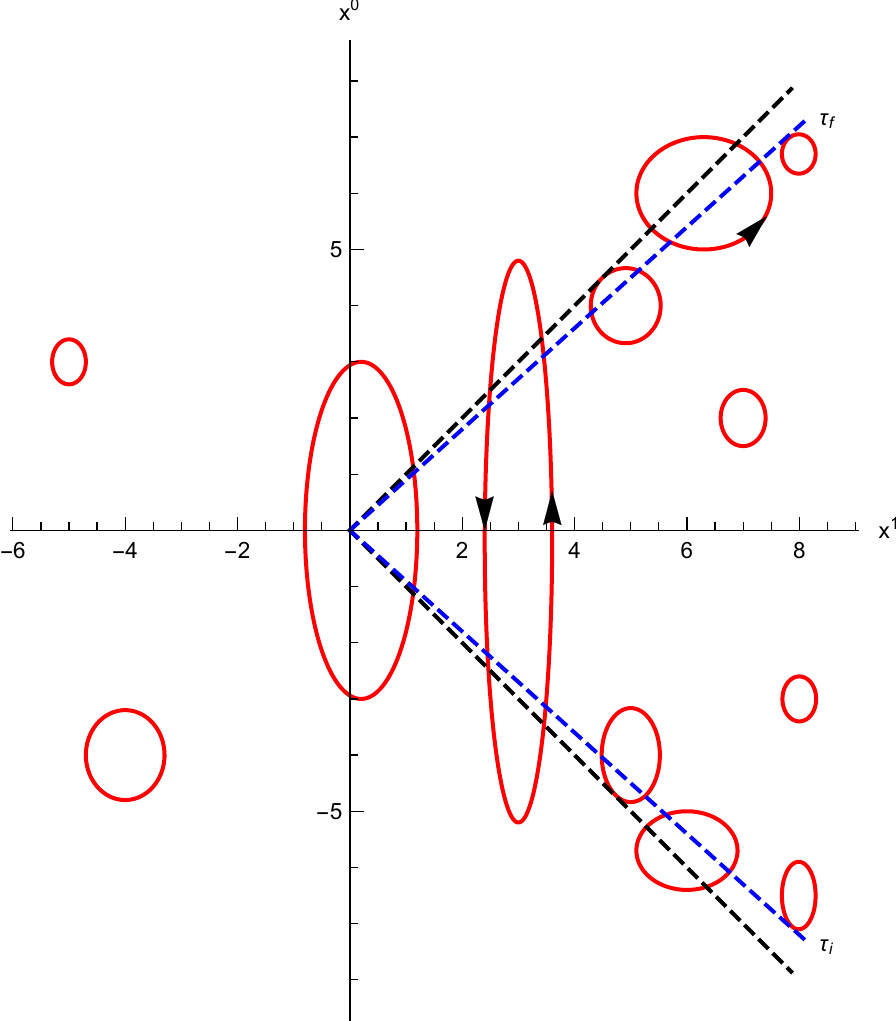}
   \caption{\sl Representation of the infinitely many  loops in Minkowski space as in Fig.\ref{fig222}, relevant to a Rindler observer for the vacuum energy calculation.}\label{fig333}
     \end{figure}

To visualize  the missing contributions we will proceed as follows. We will add to Fig.\ref{fig222}  the Rindler horizons and an initial and final space-like slices in $M_R$, as  in Fig.\ref{fig333}. They can be drown anywhere within the  wedge. Even at   $\tau_i\rightarrow -\infty$ and $\tau_f\rightarrow \infty$, which represent the horizons. These space-like slices intersect  the points where the acceleration is turned on and off.  The portion of Minkowski space the Rindler observer has access to is  bounded by the lines $\tau_i$ and $\tau_f$. On the slice $\tau_i$, a Minkowski observer turns into a Rindler observer when the acceleration is turned on. On the slice $\tau_f$, it occurs in the other way around when the acceleration is turned off.

\begin{figure}[hbt]
\centering
  \includegraphics[width=.4\textwidth]{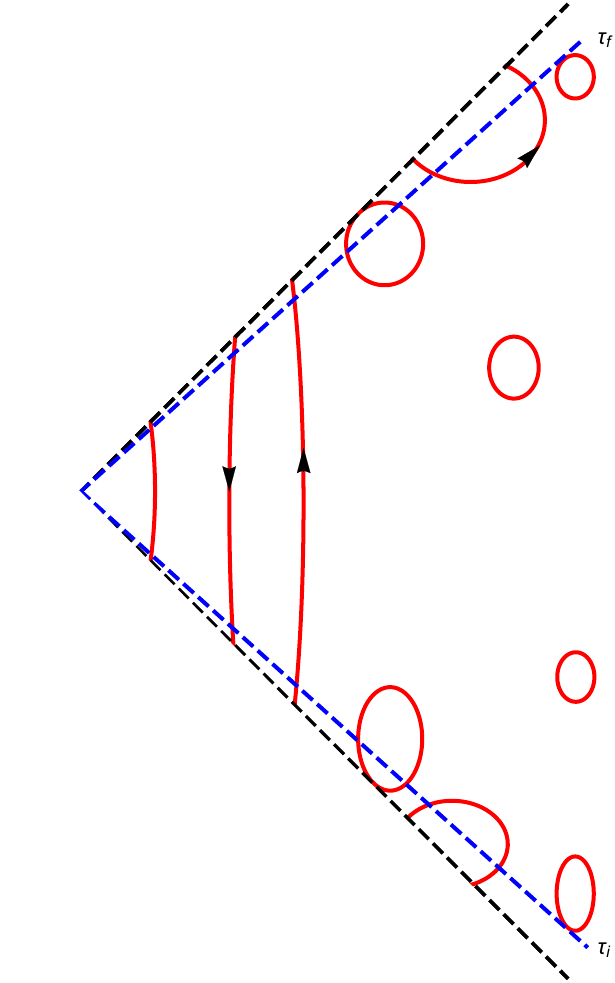}
  \caption{\sl Representation of the Rindler observer perspective of the vacuum processes. The arrows represent the flow of time in the paths.}\label{fig444}
\end{figure}

The next step is to detach  $M_R$ and its content  from the rest of the space,  Fig.\ref{fig444}. It will allow us more easily  visualize  the Rindler observer point of view.
 From Fig.\ref{fig444} we can see that in the  calculation of $E^R_{vac}$  we should take into account, in addition to $E^R_{0}$ coming from the loops,  the following open paths contributions:
 \begin{itemize}
\item All the paths going from the slice  $\tau_i$ to the slice $\tau_i$.
\item All the paths going from the slice  $\tau_f$ to the slice $\tau_f$.
\item All the paths going from the slice  $\tau_i$ to the slice $\tau_f$,  and vice versa.
\end{itemize}
Note, that from the  perspective of the Rindler observer, there are open paths that run backward in time. {\it They represent real antiparticles}.

This picture resembles\footnote{I would like to thank   I.~J.~Araya for pointing out this reference.} the one described in Chapter 3 of \cite{Susskind:2005js}. However, in this reference there is no mention to the real antiparticles in the Unruh radiation.

From Fig.\ref{fig444} we can conclude  that there are real propagating particles and antiparticles in Rindler space contributing to $E^R_{vac}$.  In fact, it is not difficult to infer this just by inspecting the arrows in Fig.\ref{fig444}. Notice that  we have not assumed the existence of any thermal state. Actually, at this stage of this derivation we do not know yet if $E^R_{vac}$ is given by a thermal distribution. We find pertinent to remark this at this point because a well known result in QFT is that any relativistic thermal state contains particles and antiparticles, this is the conventional picture. However,  here no thermality has been invoked to conclude that there are particles as well as antiparticles contributing to  $E^R_{vac}$.

\begin{figure}

\begin{minipage}[t]{7cm}
	\centering
		  \includegraphics[width=.6\textwidth]{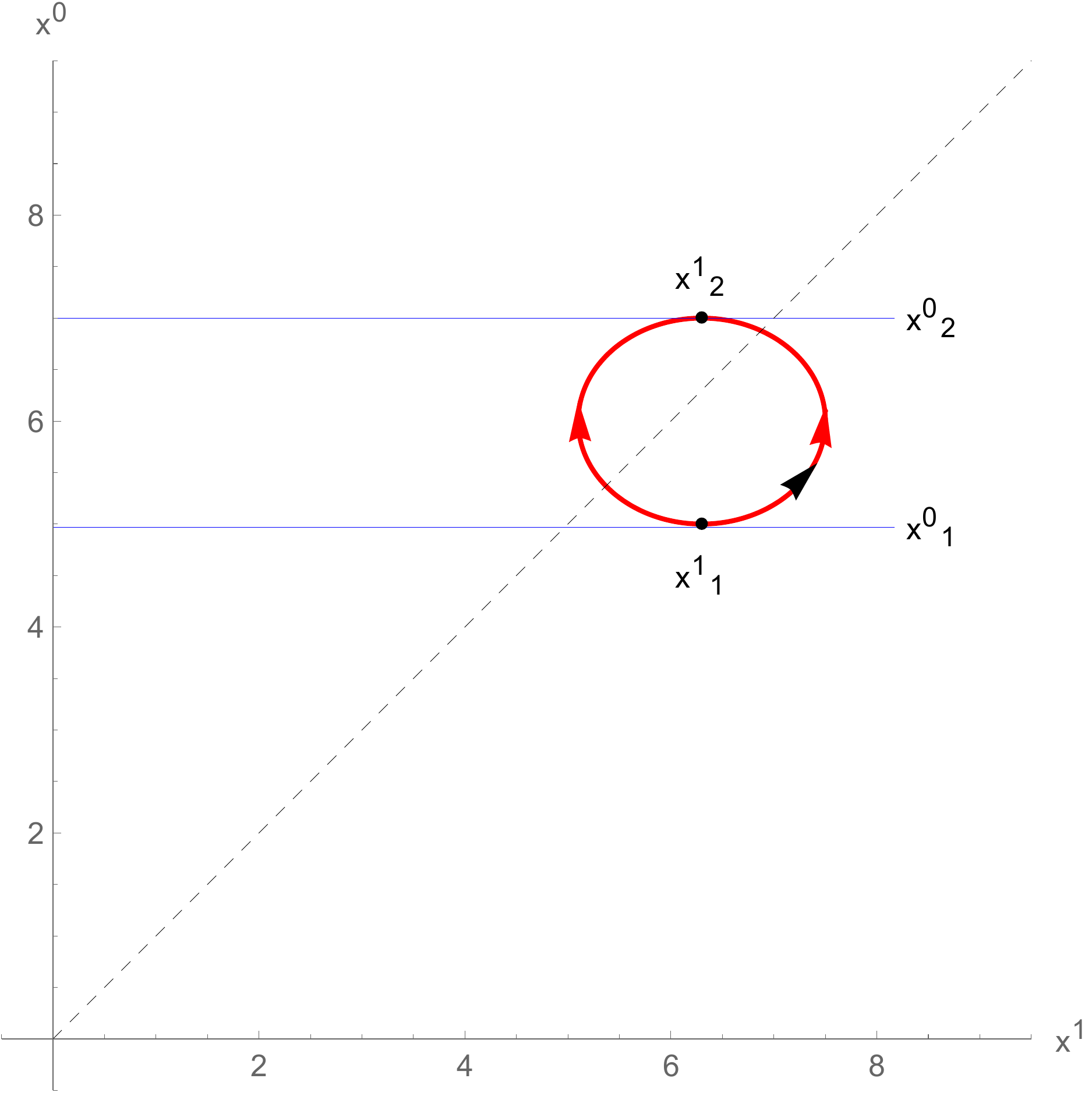}
\end{minipage}
	\hspace{1cm}
	\begin{minipage}[t]{7cm}
	\centering
		 \includegraphics[width=.6\textwidth]{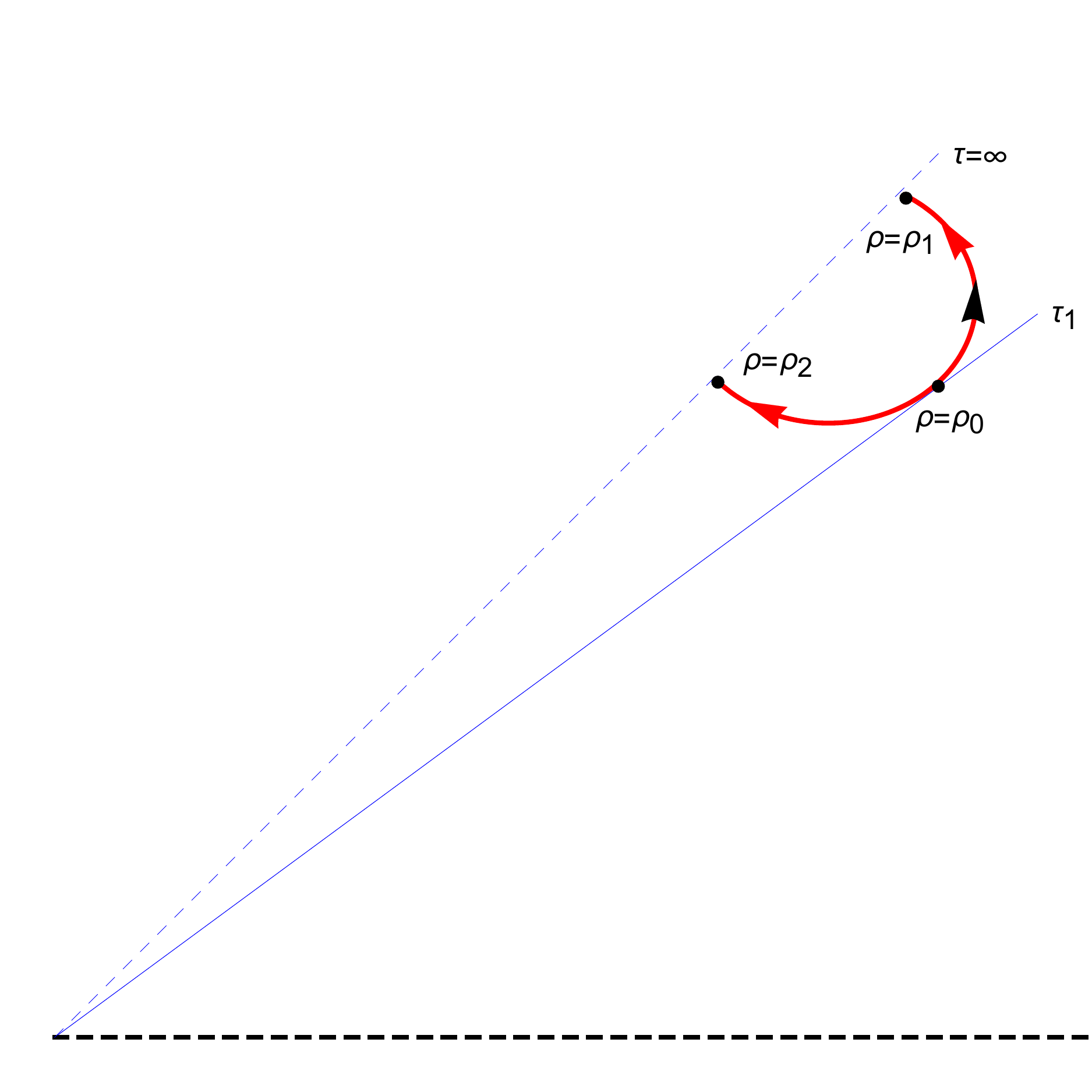}
	\end{minipage}

	\begin{minipage}[t]{7cm}
	\centering
		  \includegraphics[width=.5\textwidth]{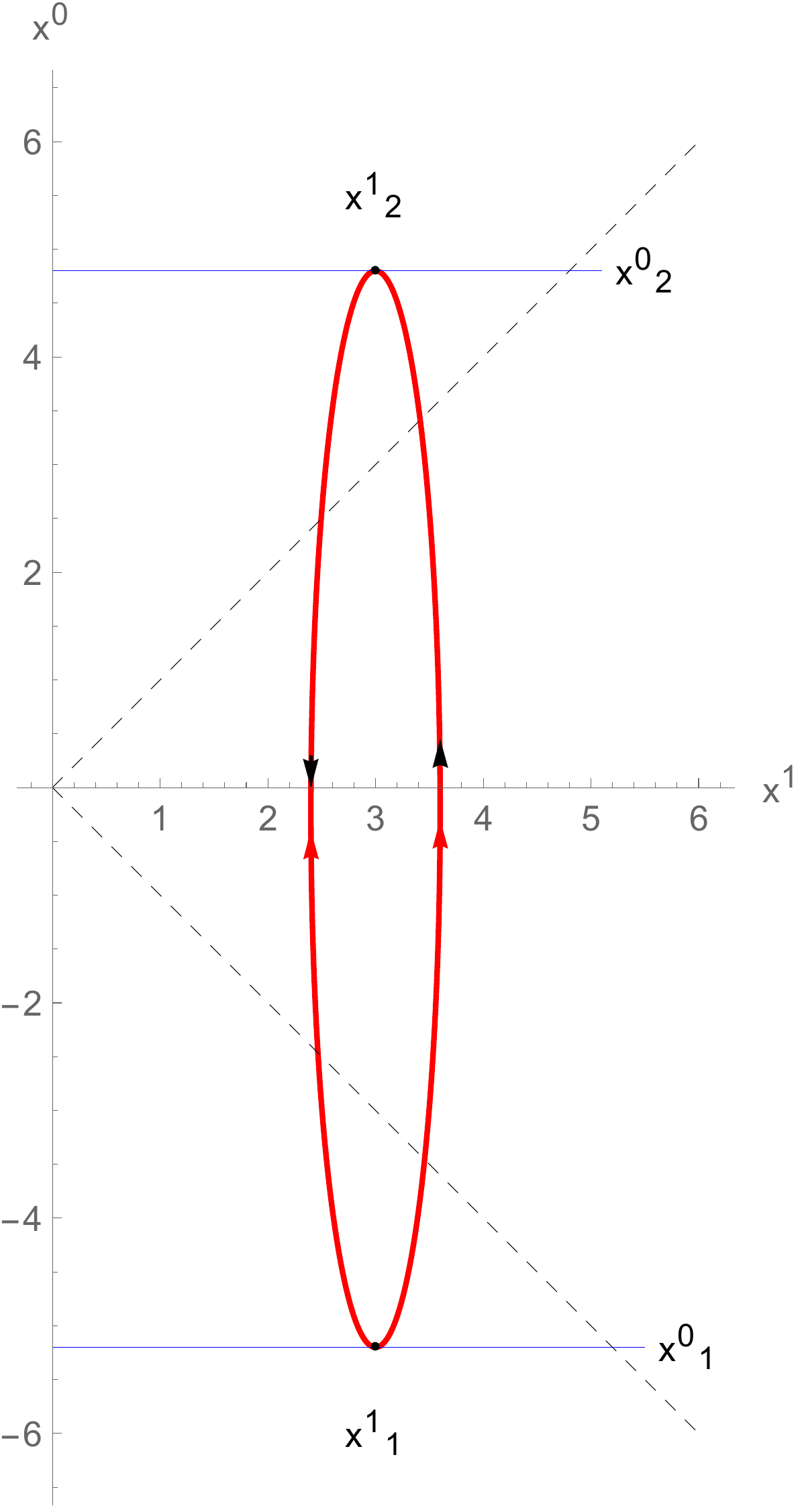}
		\caption{\sl Pictorial representation of a virtual pair  particle-antiparticle running in a loop from the Minkowski observer perspective. The black arrows represent the flow of time in the loops. The red arrows represent the direction of the propagation of  the virtual pair particle-antiparticle. In this pictures, on the one hand,  a particle creates at the point  $( x^0_1,x^1_1)$,  and  propagates forward in time to the point $( x^0_2,x^1_2)$. On the other hand, a  particle creates  at the point $( x^0_2,x^1_2)$,  and propagates backward in time to the point $( x^0_1,x^1_1)$. The path running  backward in time is interpreted as a virtual antiparticle by the Minkowski observer.}\label{mat-ant1}
\end{minipage}
	\hspace{1cm}
	\begin{minipage}[t]{7cm}
	\centering
		 \includegraphics[width=.5\textwidth]{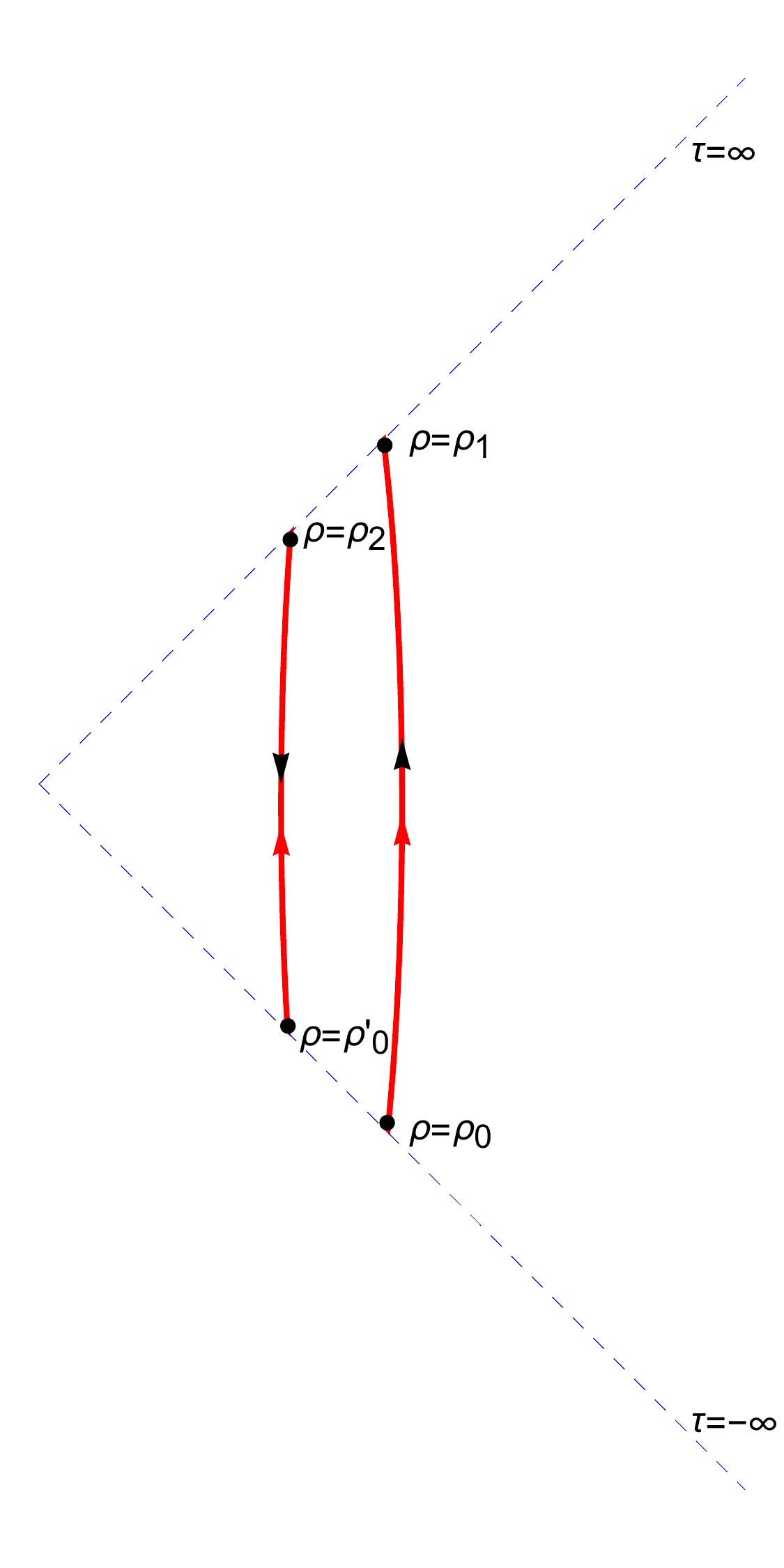}
		\caption{\sl Pictorial representation of a real pair  particle-antiparticle running in an open path (that belongs to a loop in the whole space) from the Rindler observer perspective. The black arrows represent the flow of time in the paths. The red arrows represent the direction of the propagation of the real pair particle-antiparticle. In this pictures, on the one hand,  a particle creates at the point  $( \tau_1,\rho_0)$ \big(or $( -\infty,\rho_0)$ in the second Fig.\big),  and  propagates forward in time to the point $( \infty,\rho_1)$. On the other hand, a  particle creates  at the point $( \infty,\rho_2)$,  and propagates backward in time to the point  $( \tau_1,\rho_0)$ \big(or $( -\infty,\rho'_0)$ in the second Fig.\big). The path running backward in time is seen as a real antiparticle by the Rindler observer.}\label{mat-ant2}
	\end{minipage}
\end{figure}

Let us now, explain the interpretation of each observer regarding the particles and antiparticles in correspondence with the paths and the loops. In Minkowski space,  for instance, the notion of virtual particle and antiparticle running in a loop can be visualized  as in Fig.\ref{mat-ant1}.  The same loop from the Rindler perspective looks as in  Fig \ref{mat-ant2}.
We emphasize  that  the loops (or paths)   in this paper are representatives of the infinitely many loops (or paths) passing by some given points in spacetime and they do not represent the classical picture of several particles moving in loops.

 To show that \eqref{rindlerenergyopenpaths} is the energy measured by a Rindler observer, in correspondence with the Unruh radiation,  we have to show that  $E^R_{\text{(open paths)}}$  is given by the Planck distribution with a temperature $T=\frac{1}{2\pi}$. In the next sections, after reviewing the quantization of the massless scalar field \cite{Rosabal:2018hkx}, we will show that it is indeed the case.

\section{Vacuum energy}\label{vacuumenergyseccion}

In this section we review  the massless scalar field quantization in \cite{Rosabal:2018hkx}. In this reference a non-standard quantization was presented where the field operator is subjected to the boundary conditions
\bea\label{equR1}
  \varphi_R(\tau_i,\rho,\bar{x}) & = & \varphi_M(x^0(\tau_i,\rho),x^1(\tau_i,\rho),\bar{x}), \nonumber\\
  \varphi_R(\tau_f,\rho,\bar{x}) & = & \varphi_M(x^0(\tau_f,\rho),x^1(\tau_f,\rho),\bar{x}),
\eea
where $(x^0(\tau,\rho),x^1(\tau,\rho))$, are given in \eqref{transcoor111}.
The field $\varphi_R$,  at the initial and final time is specified in Rindler space. These are the slices (see Fig.\ref{slicing}) that intersect the points where the acceleration is turned on and off. In  \cite{Araya:2017evj} similar  boundary conditions  to \eqref{equR1} were  considered, but in a different context. The operator $\varphi_M$, is known, it can be obtained from the solution of
\be
\Box \varphi_M(x^0,x^1,\bar{x})  =  0,\label{eomMink}
\ee
in the ordinary quantization in Minkowski space.

The scalar field action in the Rindler wedge $\tau_i<\tau<\tau_f$, reads
\be
S=-\frac{1}{2}\int_{\tau_i}^{\tau_f} d\tau \int_{0}^{\infty}d\rho \int_{-\infty}^{\infty} d^2x\Big[-\rho^{-1}(\partial_{\tau}\varphi_{R})^2+\rho\big((\partial_{\rho}\varphi_{R})^2+(\partial_{x^2}\varphi_{R})^2+(\partial_{x^3}\varphi_{R})^2 \big) \Big]\label{acctionRindler}.
\ee
According to the variations $\delta\varphi_{R}(\tau_i,\rho,\bar{x})=\delta\varphi_{R}(\tau_f,\rho,\bar{x})=0$, which are a consequence of \eqref{equR1}; and  $\delta\varphi_{R}(\tau,\rho,\bar{x})=0$, when $\rho\rightarrow\infty$, and $\bar{x}\rightarrow\pm \infty$, from $\delta S=0$,  we get the equation of motion
\be
\Box \varphi_R(\tau,\rho,\bar{x})  =  0\label{eomRindler}.
\ee
Here,  as in ordinary canonical quantization  we impose the equal time commutation relation,
\bea
 \big[\varphi_R(\tau,\rho,\bar{x}), \Pi_R(\tau,\rho,\bar{x}^{\prime})\big]   = \text{i}
\rho^{-1}\delta(\rho_i-\rho_f)\delta^2(\bar{x}-\bar{x}^{\prime}),
\eea
with, $\Pi_R=\rho^{-1}\partial_{\tau}\varphi_{R}$.

The general  solution of \eqref{eomMink} is
\begin{multline}\label{modeexp111}
 \varphi_M(x^0,x^1,\bar{x})  =
 \int_{-\infty}^{\infty}\frac{dk_1}{2\pi}\int_{-\infty}^{\infty}\frac{d^2k}{(2\pi)^2}\frac{1}{(2k_0)^{\frac{1}{2}}}\\
\Big(a_{(k_1,\bar{k})}\text{e}^{-\text{i}(k_0x^0+k_1x^1+\bar{k}\cdot\bar{x})}
+a^{\dagger}_{(k_1,\bar{k})}\text{e}^{\text{i}(k_0x^0+k_1x^1+\bar{k}\cdot\bar{x})}\Big).
\end{multline}
In Rindler space the general solution of \eqref{eomRindler} reads \cite{Fulling:1972md}
\be
\varphi_R(\tau,\rho,\bar{x})=\int_{0}^{\infty}d\nu\int_{-\infty}^{\infty}\frac{d^2k}{(2\pi)^2}\frac{1}{(2\nu)^{\frac{1}{2}}}
\Big(b_{(\nu,-\bar{k})}\text{e}^{-\text{i}\nu \tau}+b^{\dagger}_{(\nu,\bar{k})}\text{e}^{\text{i}\nu \tau}\Big)\psi_{\nu,\bar{k}}(\rho)\text{e}^{\text{i}\bar{k}\cdot\bar{x}},
\ee
with $k_0=(k_1^2+|\bar{k}|^2)^{\frac{1}{2}}$,  and $\psi_{\nu,\bar{k}}(\rho)$, the normalized eigenfunctions of the equation
\be
\Big(\rho^2\frac{d^2}{d\rho^2}+\rho\frac{d}{d\rho}-|\bar{k}|^2 \rho^2+\nu^2\Big)\psi_{\nu,\bar{k}}(\rho)=0,\quad \psi_{\nu,\bar{k}}(\rho)=\pi^{-1}\big(2\nu \text{sinh}(\pi \nu)\big)^{\frac{1}{2}}\text{K}_{\text{i}\nu}(|\bar{k}|\rho),\label{function}
\ee
\be
\int\limits_0^{\infty}\frac{d\rho}{\rho}\ \psi_{\nu,\kappa}(\rho)\psi_{\nu^{\prime},\kappa}(\rho)=\delta(\nu-\nu^{\prime}),
\ee
where $\text{K}_{\text{i}\nu}(|\bar{k}|\rho)$,  are the modified Bessel function of the second kind. After imposing (\ref{equR1}) we get the equations
\be
\text{e}^{-\text{i}\nu \tau_{i}}b_{(\nu,-\bar{k})}+\text{e}^{\text{i}\nu \tau_{i}}b^{\dagger}_{(\nu,\bar{k})}=
(2\nu)^{\frac{1}{2}}\int_{0}^{\infty}d\rho\int_{-\infty}^{\infty}d^2x\frac{\psi_{\nu,\bar{k}}(\rho)}{\rho}\text{e}^{-\text{i}\bar{k}\cdot\bar{x}}\varphi_{M}(\tau_{i},\rho,\bar{x}),
\ee
and
\be
\text{e}^{-\text{i}\nu \tau_{f}}b_{(\nu,-\bar{k})}+\text{e}^{\text{i}\nu \tau_{f}}b^{\dagger}_{(\nu,\bar{k})}=
(2\nu)^{\frac{1}{2}}\int_{0}^{\infty}d\rho\int_{-\infty}^{\infty}d^2x\frac{\psi_{\nu,\bar{k}}(\rho)}{\rho}\text{e}^{-\text{i}\bar{k}\cdot\bar{x}}\varphi_{M}(\tau_{f},\rho,\bar{x}),
\ee
where we are using the short hand notation
\be
  \varphi_M(\tau,\rho,\bar{x})  =  \varphi_M(x^0(\tau,\rho),x^1(\tau,\rho),\bar{x}).
\ee
Solving for $b$, and $b^{\dagger}$, we get
\bea\label{boperador}
b_{(\nu,\bar{k})} & = & -\text{i}\frac{(2\nu)^{\frac{1}{2}}}{2\text{sin}\big[\nu(\tau_f-\tau_i)\big]}\int_{0}^{\infty}d\rho\int_{-\infty}^{\infty}d^2x\frac{\psi_{\nu,\bar{k}}(\rho)}{\rho}\text{e}^{\text{i}\bar{k}\cdot\bar{x}}\Big[ \\ \nonumber
{} & {} & \text{e}^{\text{i}\nu\tau_f}\varphi_{M}(\tau_i\ ,\rho,\bar{x}) -\text{e}^{\text{i}\nu\tau_i} \varphi_{M}(\tau_f\ ,\rho,\bar{x})  \Big].
\eea

The vacuum energy $E_{vac}^R$ in Rindler space   is given by \eqref{HRindler}
\be
E_{vac}^R=\int_{-\infty}^{\infty} \frac{d^2k}{(2\pi)^2}\int_{0}^{\infty} d\nu \nu   \langle 0 | b^{\dagger}_{(\nu,\bar{k})} b_{(\nu,\bar{k})}|0\rangle\ +E^{R}_{0},\label{vacuumenergyHrindler}
\ee
where
\be
E^{R}_{0} = \frac{1}{2}\delta(0)^3 \int_{-\infty}^{\infty}d^2k  \int_{0}^{\infty}d\nu  \nu,
\ee
is the contribution of the loops  inside the Rindler space, as discussed in section \ref{OPaTE}, see also appendix \ref{appA}. This contribution most of the time is discarded.
Plugging \eqref{boperador} in \eqref{vacuumenergyHrindler} we get
\begin{multline}\label{numberofparticulas}
E_{vac}^R=\int_{-\infty}^{\infty} \frac{d^2k}{(2\pi)^2}\int_{0}^{\infty} d\nu\ \frac{\nu^2}{2 \big(\text{sin}[\nu(\tau_f-\tau_i)]\big)^2}
\\
\times\int_{0}^{\infty}d\rho\int_{-\infty}^{\infty}d^2x\frac{\psi_{\nu,\bar{k}}(\rho)}{\rho}\text{e}^{-\text{i}\bar{k}\cdot\bar{x}}
\times\int_{0}^{\infty}d\rho^{\prime}\int_{-\infty}^{\infty}d^2x^{\prime}\frac{\psi_{\nu,\bar{k}}(\rho^{\prime})}{\rho^{\prime}}\text{e}^{\text{i}\bar{k}\cdot\bar{x}^{\prime}}\Big[
\\
\langle 0 |\varphi_{M}(\tau_i,\rho,\bar{x})\varphi_{M}(\tau_i\ ,\rho^{\prime},\bar{x}^{\prime})  |0\rangle
+\langle 0 |\varphi_{M}(\tau_f,\rho,\bar{x})\varphi_{M}(\tau_f,\rho^{\prime},\bar{x}^{\prime})  |0\rangle
\\
-\text{e}^{-\text{i}\nu(\tau_f-\tau_i)}\langle 0 |\varphi_{M}(\tau_i,\rho,\bar{x})\varphi_{M}(\tau_f,\rho^{\prime},\bar{x}^{\prime})  |0\rangle
-\text{e}^{+\text{i}\nu(\tau_f-\tau_i)}\langle 0 |\varphi_{M}(\tau_f,\rho,\bar{x})\varphi_{M}(\tau_i,\rho^{\prime},\bar{x}^{\prime})  |0\rangle
\Big]\\+E_0^R.
\end{multline}

 Using the identifications \eqref{VE111} and \eqref{G111},  in terms of world line path integrals we can rewrite \eqref{numberofparticulas} as
\begin{multline}
E_{vac}^R  =  \int_{-\infty}^{\infty}\frac{d^2k}{(2\pi)^2}  \int_{0}^{\infty}d\nu\frac{\nu^2}{2 \big(\text{sin}[\nu(\tau_f-\tau_i)]\big)^2}\\
\int d^3y d^3y^{\prime}\sqrt{-g(y)}g^{\tau\tau}(y) \sqrt{-g(y^{\prime})}g^{\tau\tau}(y^{\prime})\int_0^{\infty}ds\\
 \Bigg[\Big(\psi_{\nu,\bar{k}}(\rho)\text{e}^{-\text{i}(\nu\tau_i+\bar{k}\cdot\bar{x})}\Big)\int_{y(0)\in\Sigma^{\prime}_i}^{y(1)\in\Sigma_i}[Dy^{\mu}]\text{exp}\Big[\text{i}\frac{1}{4s}\int_{0}^{1}d\sigma\ \dot{\mathbf{y}}^2\Big]\Big(\psi_{\nu,\bar{k}}(\rho^{\prime})\text{e}^{+\text{i}(\nu\tau_i+\bar{k}\cdot\bar{x^{\prime}})}\Big) \\
+\Big(\psi_{\nu,\bar{k}}(\rho)\text{e}^{-\text{i}(\nu\tau_f+\bar{k}\cdot\bar{x})}\Big)\int_{y(0)\in\Sigma^{\prime}_f}^{y(1)\in\Sigma_f}[Dy^{\mu}]\text{exp}\Big[\text{i}\frac{1}{4s}\int_{0}^{1}d\sigma\ \dot{\mathbf{y}}^2\Big]\Big(\psi_{\nu,\bar{k}}(\rho^{\prime})\text{e}^{+\text{i}(\nu\tau_f+\bar{k}\cdot\bar{x^{\prime}})}\Big) \\
-\Big(\psi_{\nu,\bar{k}}(\rho)\text{e}^{+\text{i}(\nu\tau_i+\bar{k}\cdot\bar{x})}\Big)\int_{y(0)\in\Sigma_f}^{y(1)\in\Sigma^{\prime}_i}[Dy^{\mu}]\text{exp}\Big[\text{i}\frac{1}{4s}\int_{0}^{1}d\sigma\ \dot{\mathbf{y}}^2\Big]\Big(\psi_{\nu,\bar{k}}(\rho^{\prime})\text{e}^{-\text{i}(\nu\tau_f+\bar{k}\cdot\bar{x^{\prime}})}\Big)\\
-\Big(\psi_{\nu,\bar{k}}(\rho)\text{e}^{+\text{i}(\nu\tau_f+\bar{k}\cdot\bar{x})}\Big)\int_{y(0)\in\Sigma^{\prime}_i}^{y(1)\in\Sigma_f}[Dy^{\mu}]\text{exp}\Big[\text{i}\frac{1}{4s}\int_{0}^{1}d\sigma\ \dot{\mathbf{y}}^2\Big]\Big(\psi_{\nu,\bar{k}}(\rho^{\prime})\text{e}^{-\text{i}(\nu\tau_i+\bar{k}\cdot\bar{x^{\prime}})}\Big) \\
+\text{i}\frac{1}{2}\frac{1}{V_1}\frac{1}{s}\int_{PBC}[Dy^{\mu}]\text{exp}\Big[\text{i}\frac{1}{4s}\int_{0}^{1}d\sigma\ \dot{\mathbf{y}}^2\Big] \Bigg]\ .  \label{vacpathfinal}
\end{multline}
The last term is the loop contribution $E_0^{R}$. The rest of the terms are the open path, real particle and real antiparticle, contributions as represented in Fig.\ref{fig444}, and $\Sigma$ represents a particular space-like slice.  In what follows we are going to show that \eqref{numberofparticulas} leads to  the thermal distribution.

\subsection{Vacuum energy distribution}\label{Vacuumenergydistributionseccion}

To show that \eqref{numberofparticulas} is given by the thermal distribution we will explicitly  compute the  integral
\bea\label{energiaintegral}
\int_{0}^{\infty}d\rho\int_{-\infty}^{\infty}d^2x\frac{\psi_{\nu,\bar{k}}(\rho)}{\rho}\text{e}^{-\text{i}\bar{k}\cdot\bar{x}}
\times\int_{0}^{\infty}d\rho^{\prime}\int_{-\infty}^{\infty}d^2x^{\prime}\frac{\psi_{\nu,\bar{k}}(\rho^{\prime})}{\rho^{\prime}}\text{e}^{\text{i}\bar{k}\cdot\bar{x}^{\prime}} \nonumber \\
\times\langle 0 |\varphi_{M}(\tau_i,\rho,\bar{x})\varphi_{M}(\tau_f\ ,\rho^{\prime},\bar{x}^{\prime})  |0\rangle,
\eea
this is one of the constituents of \eqref{numberofparticulas}.
Plugging (\ref{modeexp111}) and $\psi_{\nu,\bar{k}}(\rho)$,  in \eqref{energiaintegral}, and considering that the Minkowski vacuum satisfies
 \be
 a_{(p_1^{\prime},\bar{p}^{\prime})}|0\rangle=0,
 \ee
 and taking into account
\be
\big[ a_{(p_1,\bar{p})}\ , a^{\dagger}_{(p'_1,\bar{p}')}\big]=(2\pi)^3\delta(p_1-p'_1)\delta^2(\bar{p}-\bar{p}'),
\ee
we get
\bea\label{energiaintegralintermedia}
\frac{2\nu}{\pi}\text{sinh}(\pi\nu)\delta^2(0)\int_{-\infty}^{\infty}dp\frac{1}{\sqrt{p^2+|\bar{k}|^2}}\nonumber \\
\times\int_0^{\infty}d \rho\frac{\text{K}_{\text{i}\nu}(|\bar{k}|\rho)}{\rho}\text{exp}\Big[-\text{i}\rho\big(\sqrt{p^2+|\bar{k}|^2}\ \text{sinh}(\tau_i)+p \ \text{cosh}(\tau_i)\big)\Big] \nonumber\\
\times\int_0^{\infty}d \rho^{\prime}\frac{\text{K}_{\text{i}\nu}(|\bar{k}|\rho^{\prime})}{\rho^{\prime}}\text{exp}\Big[\text{i}\rho^{\prime}\big(\sqrt{p^2+|\bar{k}|^2}\ \text{sinh}(\tau_f)+p \ \text{cosh}(\tau_f)\big)\Big].
\eea
The integrals in \eqref{energiaintegralintermedia} can be reduced by means of the change of variable
\be
p=|\bar{k}|\text{sinh}(z),
\ee
to
\bea\label{interegralducida}
\frac{2\nu}{\pi}\text{sinh}(\pi\nu)\delta^2(0)\int_{-\infty}^{\infty}dz\nonumber \\
\times\int_0^{\infty}d \rho\frac{\text{K}_{\text{i}\nu}(|\bar{k}|\rho)}{\rho}\text{exp}\Big[-\text{i}\rho|\bar{k}|\text{sinh}(z+\tau_i)\Big] \nonumber\\
\times\int_0^{\infty}d \rho^{\prime}\frac{\text{K}_{\text{i}\nu}(|\bar{k}|\rho^{\prime})}{\rho^{\prime}}\text{exp}\Big[\text{i}\rho^{\prime}|\bar{k}|\text{sinh}(z+\tau_f)\Big].
\eea
Notice that we can rescale, $\rho\rightarrow |\bar{k}|^{-1}\rho$,  and $\rho' \rightarrow |\bar{k}|^{-1}\rho'$,  to  eliminate the $|\bar{k}|$ dependence in \eqref{interegralducida}. For $|\bar{k}|=0$, however,   extra care is needed.  From the asymptotic  expansion \cite{Abramowitz} of  $\text{K}_{\text{i}\nu}(|\bar{k}|\rho)$,   we can see that the integration is independent of $|\bar{k}|$ when   $|\bar{k}|\rightarrow0$. For the massive scalar field, the  calculation presented here  is similar but with $(p^2+|\bar{k}|^2)\rightarrow (p^2+|\bar{k}|^2+m^2)$, and this precaution is not needed.

The remaining  integrals  in $\rho$, and $\rho^{\prime}$, in \eqref{interegralducida}  can be straightforwardly  computed by using some  Laplace transformation table. After some algebra we have
\begin{multline}\label{casifinal}
\frac{\pi}{2\nu\text{sinh}(\pi\nu)}\delta^2(0)
\int_{-\infty}^{\infty}dz\Big(\text{e}^{\text{i}\nu(\tau_f-\tau_i)}(\text{i})^{-2\text{i}\nu}+\text{e}^{-\text{i}\nu(\tau_f-\tau_i)}(\text{i})^{2\text{i}\nu}  \\
+\text{e}^{\text{i}\nu(2 z+\tau_i+\tau_f)}+\text{e}^{-\text{i}\nu(2 z+\tau_i+\tau_f)} \Big).
\end{multline}
The two term in the second line of \eqref{casifinal} contribute with a Dirac delta function $\delta(\nu)$,  after integration in $z$.
They can be discarded because of we are removing the constant mode $\nu= 0$. After some algebra we can see that if we choose  the branches $(\text{i})^{-2\text{i}\nu}=(\text{i})^{2\text{i}\nu}=\text{e}^{-\pi \nu}$,  we obtain
\be
\frac{2\pi}{\nu}\frac{1}{\text{e}^{2\pi\nu}-1}\text{cos}\big[\nu(\tau_f-\tau_i)\big]\delta^2(0)\int_{-\infty}^{\infty}dz. \label{integralfinallll}
\ee
Plugging (\ref{integralfinallll}) in (\ref{numberofparticulas}),  we get
\be
E^{R}_{vac}= \int_{-\infty}^{\infty}\frac{d^2k}{(2\pi)^2}  \int_{0}^{\infty}\frac{d\nu}{2\pi} \frac{\nu}{\text{e}^{2\pi \nu}-1}V_3+E_0^R, \nonumber
\ee
where $V_3$, is the volume of space and we used \eqref{volumendelta}. This concludes the proof that the vacuum energy measured by a Rindler observer contains open path contributions and is given by the Planck distribution. We emphasize that  these paths represent real entangled pairs of particle-antiparticles from the Rindler perspective. Some of these pairs are confined entirely to the right Rindler wedge.  By contemplating the possibility of detecting the antiparticles in the radiation,  the quest for the Unruh radiation could be expanded.

\section{Experimental detection, an outlook}\label{experimentdet}

\begin{figure}
\centering
 \includegraphics[width=.5\textwidth]{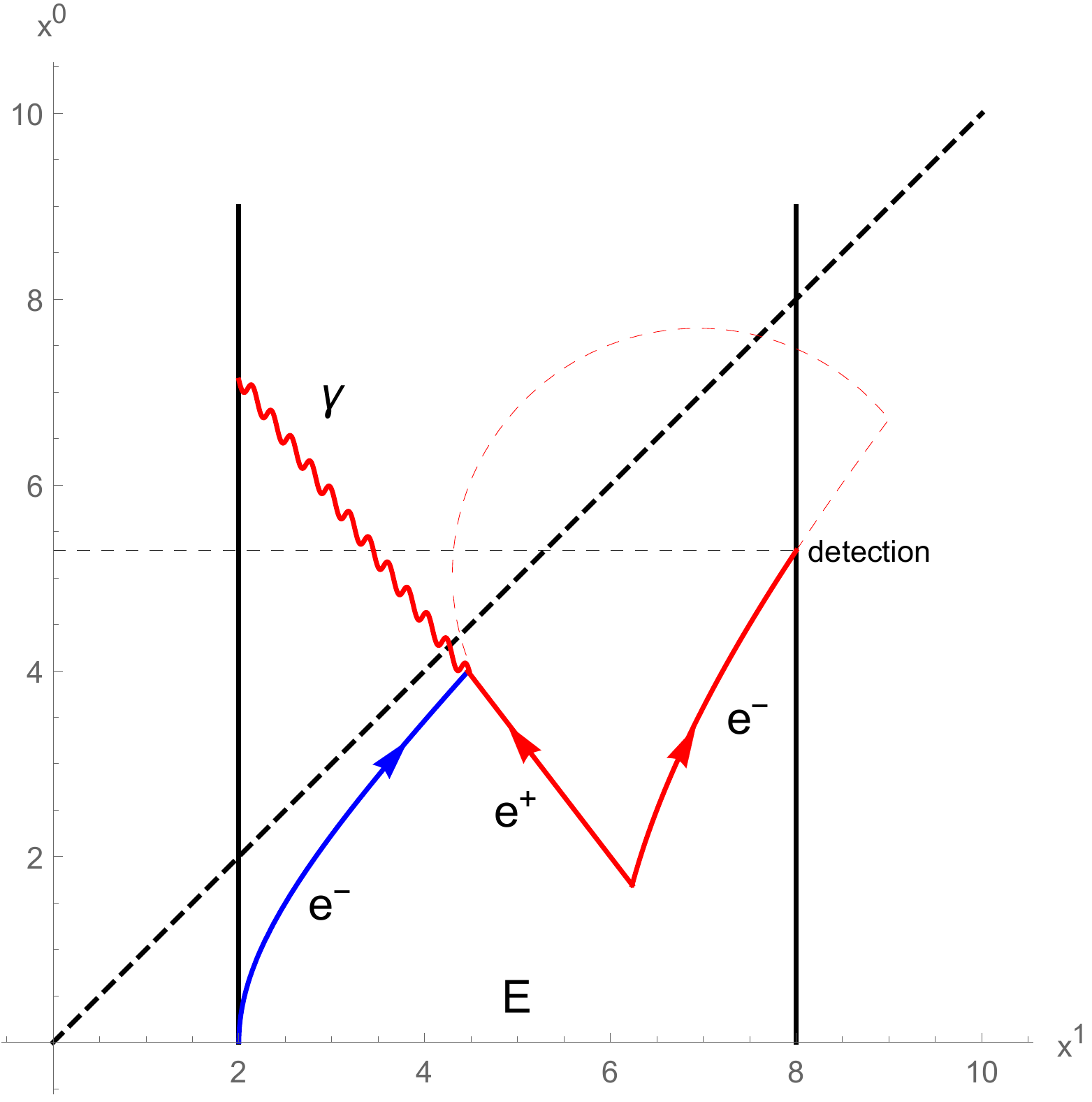}
\caption{\sl The emitter is at $x^1=2$ and the detector is at $x^1=8$ and in between a constant electric field $\text{E}$. The blue line is the ion trajectory. The red straight, curved and  wavy   lines represent  the positron, the  electron and  the photon respectively. The thin dashed black line represents  the detection time, notice that it could be smaller than the expected detection  time of the ion. The dashed red line is the path that would have completed the particle-antiparticle trajectory if the ion had not  interacted with the positron of the pair.}\label{fig6}
\end{figure}

In this section we shall describe schematically and qualitatively\footnote{A quantitative description of this experiment will be presented in \cite{Diaz_Rosabal}. In \cite{Diaz_Rosabal} the probability amplitude  for the process described in Fig.\ref{fig6} is computed in analogy with the reference \cite{Hartle:1976tp}.} a new type of experiment to detect the Unruh effect. We will  exploit the presence of antiparticles in the Unruh radiation, and the fact that they emerge out from the vacuum in entangled pairs of particle-antiparticle, as described in this work.  This is a fact that has not been used to detect the effect,  and has not been reported in the literature so far.

Over the paper we have used, for simplicity, a massless scalar field but we mentioned that the results above can be straightforwardly extended to a massive scalar field. The same machinery can also be applied to fermionic fields leading to similar conclusions about the Unruh radiation associated to these fields. The subsequent discussion will be referred to $\frac{1}{2}$ spin  charged massive particles i.e., electrons and positrons.

For an uniformly accelerated observer the trajectory in Minkowski spacetime coordinates  is
\be
(x^1)^2-(x^0)^2=\frac{1}{a^2}\label{classtraj}.
\ee
According to what we have exposed here, an observer moving along (\ref{classtraj})  will be able to interact with real antiparticles.

We will consider classical charged particles as  Rindler observers (or accelerated detectors). They could be heavy negatively charged ions of charge $q$, or any charged particle with a (almost) classical behaviour. The idea is that when accelerate a negatively charged ion  in a constant electric field,  this particle becomes a Rindler observer. As it becomes a Rindler observer it should be able to experience  the vacuum fluctuations inside the Rindler wedge as  real  particle-antiparticle pairs. If the accelerated ion meets a  positron of an  entangled  pair  they will interact by annihilation producing  a photon. After this interaction the electron in the entangled pair is released and it could be detected at some particle detector placed at some distance from the origin as represented in Fig.\ref{fig6}. {\it Detecting at least one electron would be an indication that the accelerated ion interacted  with an antiparticle from the Unruh radiation}. Of course,  increasing the number of possible events by injecting into the system a huge number of ions would increase the probability of detecting more electrons.

 The aim of this experiment is creating  a similar setup to  when Hawking radiation is emitted  \cite{Hartle:1976tp}, but in flat space. The electron detector, which is a Minkowski observer, will describe a similar situation of that observer at some constant distance from the horizon of a black hole,  as described in  \cite{Hartle:1976tp}. With some caution the work in  \cite{Hartle:1976tp}  can be interpreted as: {\it a pair particle-antiparticle creates and one falls into the black hole and is annihilated at the singularity,  allowing the other to scape}.

\section{Conclusions}

In this work we have expanded the proposal of \cite{Rosabal:2018hkx},  highlighting the presence of antiparticles in the Unruh radiation. The perspective of the Minkowski and Rindler observer regarding the vacuum processes have been presented in detail. The picture in \cite{Rosabal:2018hkx} that allows to visualize the vacuum processes has been put in correspondence with the quantization in Rindler space,  with the boundary conditions \eqref{equR1}. For this, after reviewing the quantization in the mentioned reference, we made the connection between the vacuum energy in Rindler space $E_{vac}^R$,  and the world line path integral  representation of the one loop partition function \eqref{VE111},  and the propagator \eqref{G111}. This representation is precisely what allows us to extract the interpretation of the Fig.\ref{fig444} through the result \eqref{numberofparticulas}.

Special attention has been paid to the prediction regarding the antiparticles in the Unruh radiation. At first glance this might not look like a new result.  It is well known that a thermal state contains particles and antiparticles. By  assuming  thermality and then applying  some CPT symmetry arguments one can predict the existence of antiparticles in the radiation. Notice however, that here we did not assume the existence of  any thermal state to arrive the conclusion that the Unruh radiation contains antiparticles. In fact in section \ref{OPaTE} we first arrived to the conclusion that there are antiparticles contributing to the energy measure by a Rindler observer and then in section \ref{Vacuumenergydistributionseccion} we proved that this energy is given by the Planck distribution, as expected for the Unruh radiation. Moreover,  the way of deriving our result has the advantage that we visualise each antiparticle entangled with a particle emerging both from the vacuum in the right Rindler wedge. Precisely this entanglement is what we exploit to propose the experiment. The only assumption in this work to conclude that there are antiparticles in the radiation is the existence of a Minkowski vacuum and a reliable world line path integral description of it by means of integrations over loops.

Alternatively,  we presented a new derivation of the vacuum energy from the Rindler perspective using the  Schr{\"o}dinger kernel. To our knowledge this derivation has not been presented before in the literature. In appendix \ref{appA} a detailed explanation of it can be found. From this derivation we can see, and double check, that the processes contributing to the thermal component of the  vacuum energy in Rindler space are the open path. In other words, the thermal component could be seen as a boundary contribution from the Rindler perspective.

The derivations presented in section \ref{Vacuumenergydistributionseccion} and in appendix \ref{appA} seem to contradict the essence of the Unruh effect. It is commonly believed that the Unruh effect is a peculiar feature of eternally uniformly accelerated observers, and the horizons play a determinant role in obtaining this effect.  In section \ref{Vacuumenergydistributionseccion} and in appendix \ref{appA}, however, we have shown that independently of the location of the initial and final $\tau$ slices we can get the expected thermal distribution for the energy. In other words,  a Rindler observer does not necessarily have to be an eternally uniformly accelerated observer to ``see'' a thermal bath. But, of course, when the acceleration ceases at $\tau_f$, the accelerated observer is no longer a Rindler observer and no thermal particles are detected beyond this point.

To be more concrete an explanation about the previous paragraph is needed.  The energy is a local concept, i.e., it is computed on a particular space-like slice. For instance, in Minkowski spacetime we can compute the energy at a particular $x^0$, slice \eqref{HM} and then using the time translation symmetry of the free field theory under consideration we can conclude that the energy takes the same value at any time. In Rindler space we can proceed in a similar way. We can compute the energy at any constant $\tau$, \eqref{HR} and then using the boost invariance of the theory ($\tau$-translation symmetry in the Rindler wedge) we conclude that the energy has the same value and it is independent of the particular choice of the space-like slice. In our derivation we should take care about the boundary conditions because they could break the boost symmetry at the matching slices. However, as shown these boundary conditions do not break such invariance, i.e., the final result does not depend on $\tau_i$, and $\tau_f$,  and the excepted result for the energy is obtained.

Another important point we must address is related to the Minkowski vacuum and how it is seen by a Rindler observer. Our derivation suggests that in addition to the entangled Rindler pairs of particle-antiparticle  in the Unruh state\footnote{In our approach they would correspond to the loops having a portion in the left wedge, as in \cite{Susskind:2005js}, p.41.} \cite{Unruh:1976db, Unruh:1983ms, Takagi:1986kn} there are also entangled pairs of Rindler particle-antiparticle confined to the right wedge \footnote{In our approach they would correspond to the loops having a portion in the past wedge or a portion in the future wedge.}. This result does not contradict the previous ones. The apparently mismatch is just a matter of perspective i.e., it depends upon the observer (coordinate system) we are using to make conclusions about what is real and what is virtual.

\begin{figure}[]
\centering
  \includegraphics[width=.8\textwidth]{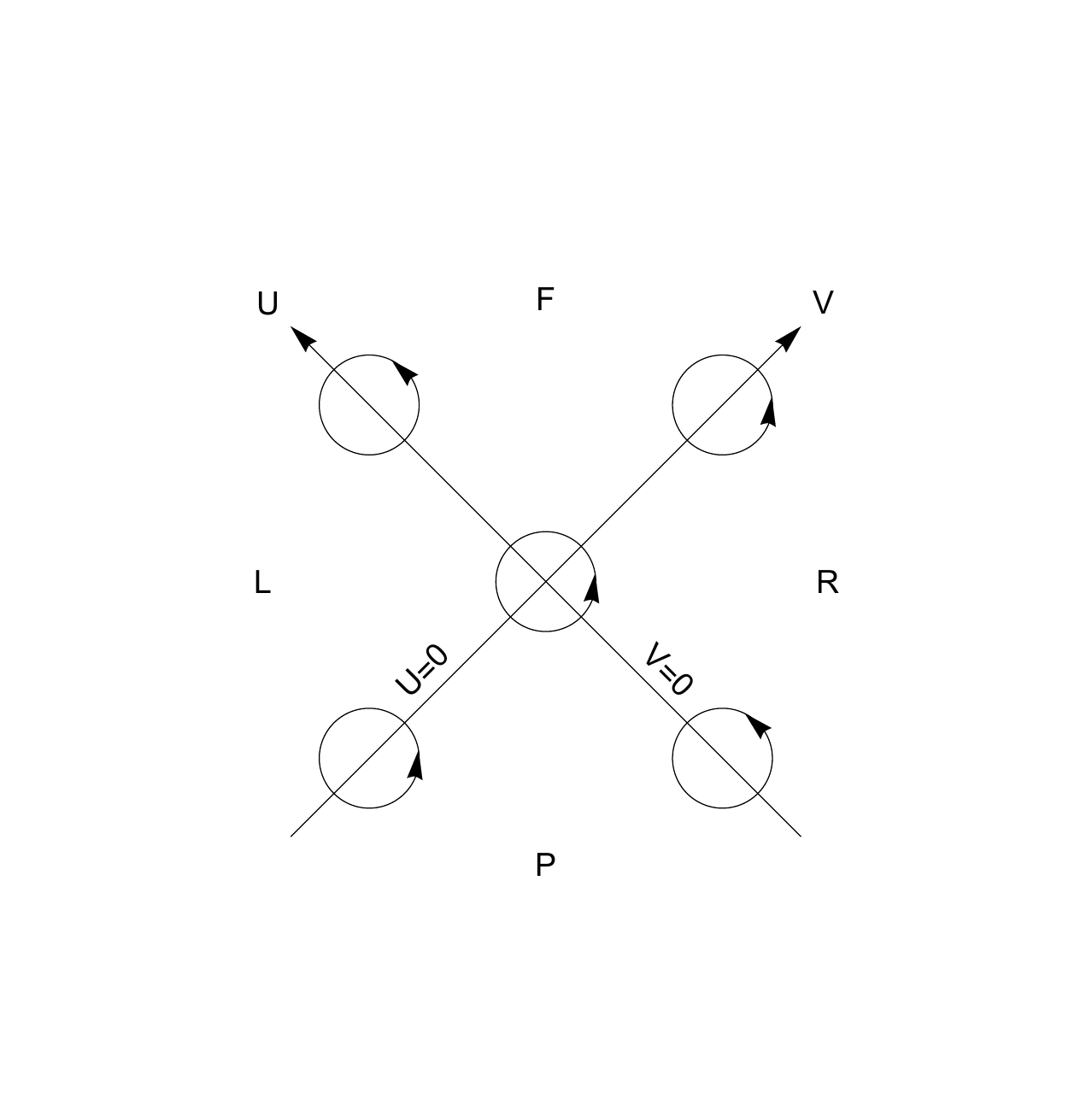}
  \caption{\sl $(U,V)$ coordinate system. R, L, P and F represent the right, left, past and future wedge respectively. The  arrows in the loops represent the flow time. $V=0$ is the past horizon while $U=0$ is the future horizon.} \label{Ustate}
\end{figure}

Let us explain  this statement. The usual picture arising from the Unruh state is that there are pairs of entangled Rindler particle-antiparticle, being the particles in the right wedge and the antiparticles in the left wedge.The derivation of this state in \cite{Unruh:1976db} uses light cone coordinates $(U, V)$, Fig.\ref{Ustate}
\bea
U & = & x^0-x^1 \nonumber,\\
V & = & x^0+x^1 \nonumber.
\eea

In this coordinates $U$, and $V$, are at the same footing i.e., either $U$, or $V$, could be regarded as time. Note also that the equation of motion is a first order partial differential equation in $U$, and $V$,
\be
\partial_{x_0}\partial_{x_0}-\partial_{x_1}\partial_{x_1}\propto \partial_{U}\partial_{V}\label{UVequ},
\ee
In this coordinates to fully determine a solution of the equation of motion, instead of specifying the field operator and its $x^0$, derivative at some constant $x^0$, we can specify the field operator at some constant $V$, ($V=0$, in \cite{Unruh:1976db} which is the past horizon), and at some constant $U$, (in \cite{Unruh:1976db} the operator has not been specified at any constant $U$ ). The relation between the Minkowski modes and, the left and right Rindler modes can be found  upon comparing the Minkowski solution and the left/right Rindler solutions at $V=0$.

Now, an  $(U, V)$ observer that regards $U$ as time (as in \cite{Unruh:1976db})  will perceive the loops intersecting the past horizon as open path, Fig.\ref{Ustate}. Notice that for $V>0$,  these paths appear as particles evolving forward in the $U$ direction (see the arrows of the loops in Fig.\ref{Ustate}). In contrast for $V<0$, the other portion of the loops evolves  backward in the $U$ direction, these can be identify with antiparticles, and we arrive to the usual interpretation of the Unruh state. What about the loops that only intersects the future horizon?

In these coordinates  the right Rindler modes can be extended through the future and past  horizon to the future and past wedges respectively. Also the left Rindler modes can be extended in a similar way. For the subsequent discussion we will focus on the right Rindler modes and their extension to the future wedge. An $(U, V)$ observer using the extended modes in the right wedge (the so-called Unruh modes) will have access to both sides of the future horizon \footnote{Note that this observer is artificial and strictly speaking is not a Rindler observer}, for $V>0$. So, the closed paths that only intersects the future horizon will look as loops and not as open path, namely virtual processes. This observer will regard this loops just as vacuum.

If instead one sticks to the $(\tau, \rho)$ coordinates, which cover only the right Rindler wedge and can not be extended beyond the horizons,  but uses the boundaries conditions \eqref{equR1} one could infer the  contributions we found. Notice that with the aim of extending the left modes only, we can use $(U, V)$ coordinates in the left wedge but $(\tau,\rho)$ coordinates in the right wedge. In this case we can extended the left modes to the future and past wedge, but we have to  use  matching conditions at $V=0$ for $U<0$ and $U=0$ for $V>0$. These conditions are compatible with the structure of the equation \eqref{UVequ}. Furthermore, they are the right conditions to fully determine the solution of \eqref{UVequ} in the whole Minkowski space. These are the conditions we have used \eqref{equR1} but in a different coordinates. It is worth to remark that in our derivation we did not make any reference to the left, past and future wedges because instead of using the artificial extended left/right Rindler modes, we used the Minkowski modes to define the boundary conditions for the right Rindler modes \eqref{equR1}. \\

In this work we have described a new kind of experiment to confirm the Unruh effect by detecting the antiparticles in the radiation.
There is a resemblance between the experiment described here and the picture presented in reference \cite{Hartle:1976tp}  by Hartle and Hawking, which explains how a black hole radiates. In this reference the world line path integral in the Schwarzschild geometry was solved by the WKB approximation. In this approximation, the solution of the classical equation of motion is a path connecting the singularity and a point at some distance from the horizon. This kind of path resembles  the (red) electron-positron trajectory in a pair  depicted in Fig.\ref{fig6}.

These similarities could be further exploited. In fact, according to what has been presented in this paper, from the perspective  of a Minkowski observer an uniformly accelerated charged particle  should radiate not only photons but also electrons  in a similar way  a black hole radiates particles, yielding perhaps to the evaporation of the Rindler observer. Moreover, taking as a guide \cite{Hartle:1976tp}, we could derive  the radiance of the accelerated charged particle\footnote{ The reader is encouraged to wait the quantitative results coming up next in \cite{Diaz_Rosabal}.}. Interestingly enough, for some set of initial and final points \cite{Diaz_Rosabal} in the Rindler wedge, under the influence of a constant electric field the solution of the classical equation of motion in Minkowski space  describes a trajectory  as the red  in Fig.\ref{fig6}.

For sure,  confirming these observations, in connection with \cite{Hartle:1976tp},  would be an step forward in our understanding, although, perhaps the origin of the Hawking radiation will remain a mystery. However,  {\it we believe} that with the current technology,  an experiment based on what has been described in section \ref{experimentdet}  could be designed,  and finally the Unruh effect could be tested.

\vspace{1cm}

\section*{Acknowledgments}

I am grateful to H. Casini,  Pablo Diaz  and Kanghoon Lee  for discussions and useful comments.

\appendix

\section{\label{appA}Quantization}

In this appendix we briefly review the quantization of a scalar field in Minkowski and  Rindler space. We compute the vacuum energy using the Hamiltonian operator and then we compare the result obtained from the one loop partition function. The one loop partition function is not explicitly computed but we used the fact that the Schr{\"o}dinger kernel  satisfies the equation
\be
 \text{i}\partial_{s}K=-\frac{1}{2}\Box K. \label{equacionkernel}
 \ee
A new derivation based on the Schr{\"o}dinger kernel and the boundary conditions \eqref{bcondikernel0} and \eqref{bcondikernel} of the Unruh effect is presented.

The action in Minkowski space is given by
\be
S=\frac{1}{2}\int_{-\infty}^{\infty} d^4x\Big[(\partial_{x^0}\varphi_{M})^2-(\partial_{x^1}\varphi_{M})^2-(\partial_{x^2}\varphi_{M})^2-(\partial_{x^3}\varphi_{M})^2 \big) \Big],
\ee
and the Hamiltonian at a particular $x^0$, slice
\be
H=\frac{1}{2}\int_{-\infty}^{\infty} d^3x\Big[(\partial_{x^0}\varphi_{M})^2+(\partial_{x^1}\varphi_{M})^2+(\partial_{x^2}\varphi_{M})^2+(\partial_{x^3}\varphi_{M})^2 \big) \Big]\label{HM}.
\ee
The equation of motion
\be
\Box_M \varphi_{M}=\partial_{x^0}\partial_{x^0}\varphi_{M}-\partial_{x^1}\partial_{x^1}\varphi_{M}-\partial_{x^2}\partial_{x^2}\varphi_{M}-\partial_{x^3}\partial_{x^3}\varphi_{M} =0, \label{eqomappedixMin}
\ee
where we take  the variations of the action  restricted  as
\be
\delta\varphi_{M}(x^0\rightarrow\pm\infty,x^1\rightarrow\pm\infty,x^2\rightarrow\pm\infty,x^3\rightarrow\pm\infty)=0,
\ee
which is a consequence of the boundary conditions
\be
\varphi_{M}(x^0\rightarrow\pm\infty,x^1\rightarrow\pm\infty,x^2\rightarrow\pm\infty,x^3\rightarrow\pm\infty)=0. \label{boundappedixMin}
\ee

The general solution of the equation of motion can be written as
\begin{multline}\label{modeMinApp}
\varphi_M(x^0,x^1,\bar{x})=\int_{-\infty}^{\infty}\frac{dk_1}{2\pi}\int_{-\infty}^{\infty}\frac{d^2k}{(2\pi)^2}\frac{1}{(2k_0)^{\frac{1}{2}}}\\
\Big(a_{(k_1,\bar{k})}\text{e}^{-\text{i}(k_0x^0+k_1x^1+\bar{k}\cdot\bar{x})}+a^{\dagger}_{(k_1,\bar{k})}\text{e}^{\text{i}(k_0x^0+k_1x^1+\bar{k}\cdot\bar{x})}\Big).
\end{multline}
Imposing the equal time canonical commutation relation
\bea\nonumber
 \big[\varphi_M(x^0,x^1,\bar{x}), \partial_{x^0}\varphi_M(x^0,x^{1\prime},\bar{x}^{\prime})\big]   = \text{i}
 \delta(x^1-x^{1 \prime})\delta^2(\bar{x}-\bar{x}^{\prime}),
\eea
we find that the operators  $a_{(k_1,\bar{k})}$ and $a^{\dagger}_{(k_1,\bar{k})}$ satisfy
\be
\big[a_{(k_1,\bar{k})},a^{\dagger}_{(k_1^{\prime},\bar{k}^{\prime})} \big]=(2\pi)^3\delta(k_1-k_1^{\prime})\delta^2(\bar{k}-\bar{k}^{\prime}).
\ee
Using the equation of motion (\ref{eqomappedixMin}) and the boundary conditions (\ref{boundappedixMin}) the Hamiltonian can be written as
\be
H^{M}=\frac{1}{2}  \int_{-\infty}^{\infty} d^3x\Big[\partial_{x^0}\varphi_{M}\partial_{x^0}\varphi_{M}-\varphi_{M} \partial_{x^0}\partial_{x^0}\varphi_{M}\Big].
\ee
After plugging (\ref{modeMinApp}) in the Hamiltonian we obtain
\be
H^{M}= \int_{-\infty}^{\infty} \frac{d^3k}{(2\pi)^3} k_0 \ a^{\dagger}_{(k_1,\bar{k})} \  a_{(k_1,\bar{k})}+E^{M}_{0},
\ee
where
\be
E^{M}_{0}=
\frac{1}{2} \delta(0)^3\int_{-\infty}^{\infty} d^3k k_0,
\ee
with $k_0= \big(k_1^2+|\bar{k}|^2\big)^{\frac{1}{2}}$.

Alternatively  $E^{M}_{0}$ can be computed by means of the Schr{\"o}dinger kernel  in Minkowski space. The solution of
 \be
 \text{i}\partial_{s}K_M(x,x^{\prime},s)=-\frac{1}{2}\Box_M K_M(x,x^{\prime},s),
 \ee
 in Minkowski space,  with the boundary condition
\be\label{kernelsolutionminBcondition}
K_M(x,x^{\prime},s)\big{|}_{s\rightarrow \pm\infty}=0,
\ee
is given by
\be\label{kernelsolutionmin}
K_M(x,x^{\prime},s)=  \int_{-\infty}^{\infty}\frac{d^4k}{(2\pi)^4}\text{e}^{\text{i}k_{\mu}(x-x^{\prime})^{\mu}} \text{e}^{\text{i}(-k_0^2+k_1^2+|\bar{k}|^2)\frac{s}{2}}.
 \ee
 To see that \eqref{kernelsolutionmin} satisfies the boundary conditions \eqref{kernelsolutionminBcondition},  we can use the integrated form of the Kernel, i.e.,
 \be\label{kernelsolutionminintegrate}
K_M(x,x^{\prime},s)= \frac{1}{(2\pi \text{i} s)^2}\text{e}^{\text{i}\frac{1}{4 s}(x-x^{\prime})_{\mu}(x-x^{\prime})^{\mu}}.
 \ee
 Using the relation
 \be
 E^{M}_{0}  =  \text{i}\frac{1}{2}\frac{1}{V_1}\int_{0}^{\infty}\frac{ds}{s}\int d^4x K_M(x,x,s) \label{pathintRindler},
 \ee
with \eqref{kernelsolutionmin} we get exactly (\ref{ecerominko}) and hence  (\ref{e0minkprimera}).

 Let us move now to the quantization  in Rindler space. The action in the Rindler patch is given by
\be
S=\frac{1}{2}\int_{\tau_i}^{\tau_f} d\tau \int_{0}^{\infty}d\rho \int_{-\infty}^{\infty} d^2x\Big[\rho^{-1}(\partial_{\tau}\varphi_{R})^2-\rho\big((\partial_{\rho}\varphi_{R})^2+(\partial_{x^2}\varphi_{R})^2+(\partial_{x^3}\varphi_{R})^2 \big) \Big],
\ee
and the Hamiltonian at a particular $\tau$, slice
\be
H^{R}=\frac{1}{2} \int_{0}^{\infty}d\rho \int_{-\infty}^{\infty} d^2x\Big[\rho^{-1}(\partial_{\tau}\varphi_{R})^2+\rho\big((\partial_{\rho}\varphi_{R})^2+(\partial_{x^2}\varphi_{R})^2+(\partial_{x^3}\varphi_{R})^2 \big) \label{HR} \Big].
\ee
The equation of motion takes the simple form
\be
\Box_R \varphi_{R} =\rho^{-1}\partial_{\tau}\partial_{\tau}\varphi_{R}-\partial_{\rho}(\rho\partial_{\rho}\varphi_{R})-\rho\big(\partial_{x^2}\partial_{x^2}\varphi_{R}+\partial_{x^3}\partial_{x^3}\varphi_{R} \big)=0, \label{eqomappedix}
\ee
where the variations of the action are restricted according to
\be
\delta\varphi_{R}(\tau_i,\rho,\bar{x})=\delta\varphi_{R}(\tau_f,\rho,\bar{x})=0,
\ee
  and
\be
\delta\varphi_{R}(\tau,\rho\rightarrow\infty,\bar{x}\rightarrow\pm\infty)=0,
\ee
which is a consequence of the boundary conditions
\bea
\varphi_R(\tau_i,\rho,\bar{x}) & = & \varphi_M(x^0(\tau_i,\rho),x^1(\tau_i,\rho),\bar{x}),  \nonumber\\
\varphi_R(\tau_f,\rho,\bar{x}) & = & \varphi_M(x^0(\tau_f,\rho),x^1(\tau_f,\rho),\bar{x}),
\eea
and
\be
\varphi_{R}(\tau,\rho\rightarrow\infty,\bar{x}\rightarrow\pm\infty) =  0. \label{boundappedix}
\ee

The general solution of the equation of motion can be written as
\be\label{soluappendix}
\varphi_R(\tau,\rho,\bar{x})=\int_{0}^{\infty}d\nu\int_{-\infty}^{\infty}\frac{d^2k}{(2\pi)^2}\frac{1}{(2\nu)^{\frac{1}{2}}}
\Big(b_{(\nu,-\bar{k})}\text{e}^{-\text{i}\nu \tau}+b^{\dagger}_{(\nu,\bar{k})}\text{e}^{\text{i}\nu \tau}\Big)\psi_{\nu,\bar{k}}(\rho)\text{e}^{\text{i}\bar{k}\cdot\bar{x}}.
\ee
Imposing the canonical commutation relation
\bea\nonumber
 \big[\varphi_R(\tau,\rho,\bar{x}), \rho^{-1}\partial_{\tau}\varphi_{R}(\tau,\rho^{\prime},\bar{x}^{\prime})\big]   = \text{i}
  \rho^{-1}\delta(\rho_i-\rho_f)\delta^2(\bar{x}-\bar{x}^{\prime}),
\eea
we find that the operators  $b_{(\nu,-\bar{k})}$, and $b^{\dagger}_{(\nu,\bar{k})}$, satisfy
\be
\big[b_{(\nu,-\bar{k})}, b^{\dagger}_{(\nu^{\prime},\bar{k}^{\prime})} \big]=(2\pi)^2\delta(\nu-\nu^{\prime})\delta^2(\bar{k}+\bar{k}^{\prime}). \label{comuinrindler}
\ee
Using the equation of motion (\ref{eqomappedix}) and the boundary conditions (\ref{boundappedix}) the Hamiltonian can be written as
\be
H^{R}=\frac{1}{2} \int_{0}^{\infty}\frac{d\rho}{\rho} \int_{-\infty}^{\infty} d^2x\Big[\partial_{\tau}\varphi_{R}\partial_{\tau}\varphi_{R}-\varphi_{R} \partial_{\tau}\partial_{\tau}\varphi_{R}\Big].
\ee
Plugging (\ref{soluappendix}) in the Hamiltonian and using (\ref{comuinrindler})  we get the familiar form
\be
H^{R}= \int_{-\infty}^{\infty} \frac{d^2k}{(2\pi)^2} \int_{0}^{\infty}d\nu\nu b^{\dagger}_{(\nu,\bar{k})} \  b_{(\nu,\bar{k})}+E^{R}_{0}\label{HRindler},
\ee
where
\be
E^{R}_{0}=
\frac{1}{2} \delta(0)^3\int_{-\infty}^{\infty} d^2k \int_{0}^{\infty}d\nu\nu.
\ee
The vacuum energy can  be computed as in (\ref{vacuumenergyHrindler}).

Alternatively we can use the fact that we know how to solve the equation
 \be
 \text{i}\partial_{s}K_R(y,y^{\prime},s)=-\frac{1}{2}\hat{O} K_R(y,y^{\prime},s),
 \ee
in Rindler space, with
\be
\Hat{O}=\partial_{\tau}\partial_{\tau}-\rho\partial_{\rho}(\rho\partial_{\rho})-\rho^2\delta^{ab}\partial_a\partial_b,
\ee
and compute the vacuum energy by means of \eqref{pathintRindler0}
\bea
 E^{R}_{vac}  & = & \text{i}\frac{1}{2}\frac{1}{V_1}\int_{0}^{\infty}\frac{ds}{s}\int_{\tau_i}^{\tau_f}d\tau \int_0^{\infty} \frac{d\rho}{\rho}\int d^2\bar{y} K_R(y,y,s), \label{trazakernel}
 \eea
 where $y=(\tau,\rho,\bar{y})=(\tau,\rho,y^2,y^3)$, the metric is (\ref{metric1}) and $V_1=\int_{\tau_i}^{\tau_f}d\tau$. Notice that
\be
\hat{O}\neq\Box_R.
\ee
A nice geometrical interpretation can be given to $K_R$ and $\hat{O}$, see \cite{Emparan:1994qa}, on a hyperbolic space. However, for our purpose it is enough to work with   $K_R$ as defined on Rindler space.

Let us now computed $E^{R}_{vac} $ using (\ref{trazakernel}). The naive choice of the  solution of the equation (\ref{equacionkernel}),
 \be
K_R(y,y^{\prime},s)= \int_{-\infty}^{\infty}\frac{d\nu_0}{2\pi} \int_{0}^{\infty}d\nu \int_{-\infty}^{\infty}\frac{d^2k}{(2\pi)^2}\text{e}^{\text{i}\nu_0(\tau-\tau^{\prime})} \psi_{\nu,\bar{k}}(\rho)\psi_{\nu,\bar{k}}(\rho^{\prime})\text{e}^{\text{i}\bar{k}\cdot(\bar{x}-\bar{x}^{\prime})} \text{e}^{\text{i}(-\nu_0^2+\nu^2)\frac{s}{2}},
 \ee
leads to $E^{R}_{0}$, as we described in section \ref{vacproseccion}. To get the thermal contribution of the vacuum energy we have to take into account that in the overlap between the Rindler patch and Minkowski space  the kernel are the same. It means that from the Rindler perspective the solution of (\ref{equacionkernel}) has to satisfy,  in addition to
\be
K_R(y,y^{\prime},s)\big{|}_{s\rightarrow \pm\infty}=0,\label{bcondikernel0}
\ee
 the boundary conditions
\bea \label{bcondikernel}
K_R(y,y^{\prime},s)\big{|}_{\tau_i}& = & K_M(x(y),x^{\prime}(y^{\prime}),s)\big{|}_{\tau_i}, \nonumber\\
K_R(y,y^{\prime},s)\big{|}_{\tau_f} & = & K_M(x(y),x^{\prime}(y^{\prime}),s)\big{|}_{\tau_f},
\eea
where
\[ x(y)=   \left\{
\begin{array}{lll}
      x^0(\tau,\rho) & = & \rho \ \text{sinh}(\tau)\\
      x^1(\tau,\rho) & = & \rho \ \text{cosh}(\tau)\\
      \ \ \ \ \ \ \ \ \bar{x} & = &   \bar{x}
      \end{array}
\right. \]
and $K_M(x,x^{\prime},s)$, is given in \eqref{kernelsolutionmin}.

To solve (\ref{equacionkernel}) with (\ref{bcondikernel}) we proceed in a similar way as in the section \ref{vacuumenergyseccion}.
We write the general solution as
 \bea
K_R(y,y^{\prime},s)= \int_{0}^{\infty}\frac{d\nu_0}{2\pi} \int_{0}^{\infty}d\nu \int_{-\infty}^{\infty}\frac{d^2k}{(2\pi)^2}\Big(\alpha \ \text{e}^{-\text{i}\nu_0(\tau-\tau^{\prime})}+\beta \  \text{e}^{\text{i}\nu_0(\tau-\tau^{\prime})}\Big) \\ \nonumber
 \psi_{\nu,\bar{k}}(\rho)\psi_{\nu,\bar{k}}(\rho^{\prime})\text{e}^{\text{i}\bar{k}\cdot(\bar{x}-\bar{x}^{\prime})} \text{e}^{\text{i}(-\nu_0^2+\nu^2)\frac{s}{2}}.
 \eea
Where $\alpha$ and $\beta$ are two constants that we determine upon imposing (\ref{bcondikernel}). After a very long algebra\footnote{One encounters similar integrals as in section \ref{vacuumenergyseccion}.} one finds that  these two constants  do not depend on $\tau_i$, and $\tau_f$,
\be
\alpha=\beta=\text{coth}(\pi \nu).
\ee
So,  the proper Schr{\"o}dinger kernel in Rindler space is
\bea
K_R(y,y^{\prime},s)= \int_{-\infty}^{\infty}\frac{d\nu_0}{2\pi} \int_{0}^{\infty}d\nu \int_{-\infty}^{\infty}\frac{d^2k}{(2\pi)^2}\text{coth}(\pi \nu)\\ \nonumber
\text{e}^{\text{i}\nu_0(\tau-\tau^{\prime})} \psi_{\nu,\bar{k}}(\rho)\psi_{\nu,\bar{k}}(\rho^{\prime})\text{e}^{\text{i}\bar{k}\cdot(\bar{x}-\bar{x}^{\prime})} \text{e}^{\text{i}(-\nu_0^2+\nu^2)\frac{s}{2}}.
 \eea
This kernel can be written as
\bea \label{properkernel}
K_R(y,y^{\prime},s)= \int_{-\infty}^{\infty}\frac{d\nu_0}{2\pi} \int_{0}^{\infty}d\nu \int_{-\infty}^{\infty}\frac{d^2k}{(2\pi)^2}\Big[\frac{2}{\text{e}^{2\pi \nu}-1}+1\Big]\\ \nonumber
\text{e}^{\text{i}\nu_0(\tau-\tau^{\prime})} \psi_{\nu,\bar{k}}(\rho)\psi_{\nu,\bar{k}}(\rho^{\prime})\text{e}^{\text{i}\bar{k}\cdot(\bar{x}-\bar{x}^{\prime})} \text{e}^{\text{i}(-\nu_0^2+\nu^2)\frac{s}{2}}.
 \eea

It is not difficult to compute the vacuum energy from (\ref{properkernel}). Plugging it in (\ref{trazakernel}), taking into account the orthogonality  of the functions  $\psi_{\nu,\bar{k}}(\rho)$,
\be
\int\limits_0^{\infty}\frac{d\rho}{\rho}\ \psi_{\nu,\kappa}(\rho)\psi_{\nu^{\prime},\kappa}(\rho)=\delta(\nu-\nu^{\prime}),
\ee
 and \eqref{volumendelta}, and  (\ref{importantidentity}),  we get
\be\label{finalresult}
E_{vac}^R=  \int_{-\infty}^{\infty}\frac{d^2k}{(2\pi)^2}\int_{0}^{\infty}\frac{d\nu}{2\pi}\frac{\nu}{\text{e}^{2\pi \nu}-1}V_3+E^{R}_{0},
\ee
where the first term in \eqref{finalresult} comes from the first term in the bracket in \eqref{properkernel} and $E^{R}_{0}$ comes from the second term in the bracket in \eqref{properkernel}. We have already shown in section \ref{vacuumenergyseccion}  that the thermal contribution to the vacuum energy (\ref{finalresult}) comes from the open paths starting and ending at the horizons or in other words boundaries contributions, as shown from this derivation.

\end{document}